\documentclass[a4paper,12pt]{article}

\usepackage{graphicx}
\usepackage{mathptmx}
\usepackage[authoryear]{natbib}
\usepackage{longtable}
\usepackage{afterpage}
\usepackage{fancyhdr}
\usepackage{enumitem}
\usepackage{footmisc}
\usepackage[top=3cm, bottom=2.5cm, left=2.1cm, right=2.1cm]{geometry}

\setlist{nolistsep}
\def\institute#1{\gdef\@institute{#1}}

\begin{document}

\pagestyle{fancy} 
\lhead{Hilton et al.}
\rhead{Report of IAU Working Group on Access to Ephemerides}

\title{Report of the IAU Commission 4 Working Group on Standardizing Access to Ephemerides and File Format Specification}

\author{James~L.~Hilton\footnotemark[1],
        Charles~Acton\footnotemark[2],
        Jean\mbox{-}Eudes~Arlot\footnotemark[3],
        Steven~A.~Bell\footnotemark[4],\\
        Nicole~Capitaine\footnotemark[5],
        Agn\`{e}s~Fienga\footnotemark[6],
        William~M.~Folkner\footnotemark[2],
        Micka\"{e}l~Gastineau\footnotemark[3],\\
        Dmitry~Pavlov\footnotemark[8],
        Elena~V.~Pitjeva\footnotemark[8],
        Vladimir~I.~Skripnichenko\footnotemark[8],\\
        \& Patrick~Wallace\footnotemark[9]}
\renewcommand{\thefootnote}{\fnsymbol{footnote}}
\footnotetext[1]{U.S. Naval Observatory, Astronomical Applications Dept., 3450 Massachusetts Ave. NW, Washington, DC 20392, USA. Tel.: +1-541-324-9432, email: james.hilton@usno.navy.mil}
\footnotetext[2]{Jet Propulsion Laboratory, California Institute of Technology}
\footnotetext[3]{Observatoire de Paris IMCCE}
\footnotetext[4]{H.M. Nautical Almanac Office, UK Hydrographic Office}
\footnotetext[5]{Observatoire de Paris SYRTE}
\footnotetext[6]{Observatoire de C\^{o}te d'Azur}
\footnotetext[3]{Observatoire de Paris IMCCE}
\footnotetext[8]{Institute of Applied Astronomy}
\footnotetext[9]{RAL Space (retired)}

\maketitle

\begin{abstract}

The IAU Commission 4 Working Group on Standardizing Access to Ephemerides recommends the use of the Spacecraft and Planet Kernel (SPK) format as a standard format for the position ephemerides of planets and other natural solar system bodies, and the use of the Planetary Constants Kernel (PCK) format for the orientation of these bodies. It further recommends that other supporting data be stored in a text PCK. These formats were developed for use by the SPICE Toolkit by the Navigation and Ancillary Information Facility of NASA's Jet Propulsion Laboratory (JPL). The CALCEPH library developed by the Institut de m\'{e}canique c\'{e}leste de calcul des \'{e}ph\'{e}m\'{e}rides (IMCCE) is also able to make use of these files. High accuracy ephemerides available in files conforming to the SPK and PCK formats include: the {\em Development Ephemerides} (DE) from JPL, {\em Int\'{e}grateur Num\'{e}rique Plan\'{e}taire de l'Observatoire de Paris} (INPOP) from IMCCE, and the {\em Ephemerides Planets and the Moon} (EPM), developed by the Institute for Applied Astronomy (IAA). The bulk of this report is a description of the portion of PCK and SPK formats required for these ephemerides. New SPK and PCK data types, both called Type 20: Chebyshev (Velocity Only), have been added. Other changes to the specification are (i)~a new object identification number for coordinate time ephemerides and (ii)~a set of three new data types that use the TCB rather than the TDB time scale for the ephemerides, but are otherwise identical to their TDB versions.
 \end{abstract}
 
 \clearpage
 \tableofcontents
 \renewcommand{\thefootnote}{\arabic{footnote}}
 \vspace{12pt}
 
 \section*{Recommendation}
 \addcontentsline{toc}{section}{Recommendation}
 
To provide a uniform format for the position ephemerides of planets and other natural solar system bodies the International Astronomical Union (IAU) Commission 4: Ephemerides Working Group on Standardizing Access to Ephemerides recommends: \begin{enumerate}
\item The use of the Spacecraft and Planet Kernel (SPK) format \citep{Bachman2010} for the positions of solar system bodies.
\item Supporting data on the ephemerides and simple body orientation ephemerides be stored in text Planetary Constants Kernel (PCK) format \citep{WrightActon2013}.
\item The use of the binary PCK format for the orientation of a body when it is too complex to be represented using text PCK \citep{WrightActon2013}.
\end{enumerate}

\section{Introduction}

These file formats were  developed by the Navigation and Ancillary Information  Facility (NAIF) of NASA's Jet Propulsion Laboratory (JPL) as a portion of the  SPICE space geometry information system. The SPICE documentation refers to PCK and SPK data files as kernels; the same terminology will be followed here.

{\bf Most users will want to use either the SPICE Toolkit or CALCEPH, developed by the Institut de m\'{e}canique c\'{e}leste de calcul des \'{e}ph\'{e}m\'{e}rides (IMCCE), to access ephemerides stored in these formats.}

SPICE is an information system designed to assist in planning and interpreting scientific observations from space-based instruments. SPICE is also widely used in engineering tasks associated with planetary missions. The SPICE system includes a large suite of software that may be incorporated into application programs to read SPICE kernels and, using those data, compute derived observation geometry, such as altitude, latitude/longitude, and lighting angles. SPICE data and software may be used within many different computing environments. The software is available in FORTRAN~77, C, IDL and MATLAB from the NAIF web site \citep{NAIF2013}.

In addition to SPICE, CALCEPH, starting with version 2.0, has the ability to read text PCKs, binary PCKs, and SPKs. CALCEPH was developed primarily to read IMCCE's {\em Int\'{e}grateur Num\'{e}rique Plan\'{e}taire de l'Observatoire de Paris} (INPOP) planetary ephemerides. It is a library with interfaces to allow it to be linked to programs written in C, FORTRAN~77, and Fortran~90/95/2003. It is available at the INPOP web site \citep{CALCEPH2013} and the IAU Commission 4: Ephemerides web site \citep{IAUC4}.

At least three high-accuracy ephemerides, the JPL's {\em Development Ephemerides} (DE), the IMCCE's INPOP, and the Institute of Applied Astronomy's {\em Ephemerides of Planets and the Moon} (EPM) \citep{EPM2013} have versions available in the SPK/PCK formats.

The SPICE Toolkit contains procedures for both writing and reading SPK and PCK files. But some users, such as ephemeris developers, may want to access the ephemeris files directly or construct ephemeris files in these formats using their own software. Detailed specification of those portions of the PCK and SPK formats needed for the ephemerides of solar system bodies are required to meet this objective. This specification forms the bulk of this report.

The SPK format is designed to contain ephemerides of multiple bodies. These ephemerides may be stored using multiple data types\footnote{A data type, in this context, is a method of representing an ephemeris, such as Chebyshev polynomials or orbital elements.}. Only a few of these types are used for the storage of the position ephemerides of natural solar system bodies. The SPK format is similar to the binary PCK format \citep{Bachman2010}. The binary PCK format may, among other things, be used to store ephemerides of the orientation of bodies (such as a lunar orientation ephemeris), which are too  complex to be stored using the text PCK format. The specification for the storage of complex body orientation ephemerides is included in \S\ref{PCK}.

The SPK and binary PCK formats are based in turn on the Double Precision Array File (DAF) architecture \citep{Wright2013} to store the double precision (64-bit real number) arrays of ephemeris data. That part of the specification of the DAF architecture required to write SPKs and binary PCKs is given in \S\ref{DAF}.

The information required to specify a binary PCK or SPK fully for a natural solar system body is spread throughout the SPICE documentation, available both within the SPICE Toolkit and on the NAIF web site \citet{NAIF2013}. The main repositories of the specification of the SPK and binary PCK formats are the {\em DAF Required Reading}, the {\em SPK Required Reading}, the {\em PCK Required Reading}, and the SPICE source code files for the individual SPK and binary PCK data types. Additional data on SPKs and PCKs and how they are used and accessed with the SPICE Toolkit are available in a set of tutorial slides, also available at the NAIF web site \citet{NAIF2013b}\footnote{The tutorials of interest for this paper are: 19:~SPK~(19\_spk.pdf), 20:~PCK~(20\_pck.pdf), 25:~Lunar-Earth~PCK-FK~(25\_lunar-earth\_pck-fk.pdf), and 41:~Making an SPK~(41\_making\_an\_spk.pdf).}. The purpose of the bulk of this report is to bring together in one place that portion of the format specification required for storing an ephemeris for a natural solar system body.

Supporting data cover a wide variety of parameters and other information. Among other things, these data may include the orientation of the body as a function of time\footnote{The orientation of a few bodies, notably the Earth and the Moon, is too complex to be reasonably stored in text PCK format. For these exceptions there is the binary PCK format discussed in \S\ref{PCK}}, the parameters used in constructing the ephemerides, the values of the parameters, whether they were fixed or solved for as a part of the solution, the units used, and the initial conditions. To accommodate this wide variety of data and values, the working group recommends storing these supporting data in a text PCK. The specification of text PCK is described in \S\ref{TextPCK}.

To accommodate the requirements of the wider community, NAIF has agreed to make a number of additions to the set of SPK kernel types and to make adjustments to the SPICE Toolkit and its documentation. These changes are outlined in \S\ref{SPKChanges}. Next, \S\ref{CoordTimeEphem} describes the use of coordinate time ephemerides in SPK kernels, \S\ref{DAF} describes that portion of the DAF architecture required to understand SPKs and binary PCKs, \S\ref{SPK} describes that part of the SPK format required to store orientation ephemerides and to understand the organization of SPKs, \S\ref{PCKKernels} discusses PCKs, \S\ref{TextPCK} describes the text PCK format for storing supporting data, and \S\ref{PCK} describes the binary PCK format for storing binary Chebyshev polynomials for complex orientation ephemerides such as the Earth and the Moon.

The six SPK data types, listed in Table~\ref{SPKPCKDataTypes}, for ephemerides of natural solar system bodies are described in the sections given in that table. Types 2, 3, and 20 differ from types 102, 103, and 120 only in the fact that the independent variable in the former three is the TDB time scale while it is the TCB time scale for the latter three. The formats of the four binary PCK types, tabulated in Table~\ref{SPKPCKDataTypes}, are used to store body orientation ephemerides. The difference between the SPK types and their binary PCK counterparts is in how SPICE interprets the results of their evaluation.

\begin{table}
\begin{center}
\caption{Double precision kernel data types of interest.\label{SPKPCKDataTypes}}
\begin{tabular}{l r l c}
\hline\hline
\multicolumn{1}{c}{Kernel} & \multicolumn{1}{c}{Type} & \multicolumn{1}{c}{Description} & Section\\
  & \multicolumn{1}{c}{Number} &  & Discussed\\
\hline
SPK &   2 & Chebyshev (Position Only)              & \S\ref{Type2}\\
SPK &   3 & Chebyshev (Position and Velocity)      & \S\ref{Type3}\\
SPK &  20 & Chebyshev (Velocity Only)              & \S\ref{Type20}\\
SPK & 102 & Chebyshev (TCB: Position Only)         & \S\ref{Type2}\\
SPK & 103 & Chebyshev (TCB: Position and Velocity) & \S\ref{Type3}\\
SPK & 120 & Chebyshev (TCB: Velocity Only)         & \S\ref{Type20}\\
PCK &   2 & Chebyshev (Angles)                     & \S\ref{PCKType2}\\
PCK & 102 & Chebyshev (TCB: Angles)                & \S\ref{PCKType2}\\
PCK &  20 & Chebyshev (Angle Rates)          & \S\ref{PCKType20}\\
PCK & 120 & Chebyshev (TCB: Angle Rates)     & \S\ref{PCKType20}\\
\hline
\end{tabular}
\end{center}
\end{table}

\section{Changes Made to the binary SPK and PCK Formats}\label{SPKChanges}

To accommodate the requirements of the wider community NAIF has agreed to make additions to the SPK and PCK types and adjustments to the SPICE Toolkit and its documentation. These changes are:
\begin{enumerate}
 
\item New SPK and PCK data types have been added. The SPK data type, Type 20:  Chebyshev (Velocity Only), is described in \S\ref{Type20}, and the PCK data type, Type 20: Chebyshev (Angle Rates), is described in \S\ref{PCKType20}.

\item The data types beginning with 101 have been reserved for ephemerides where the time argument uses the TCB rather than TDB time scale. Currently, this designation is applied to SPK types 102, 103, and 120 and binary PCK types 102 and 120. These data types differ from SPK types 2, 3, and 20 and PCK types 2 and 20 only in the detail that the time argument used to extract data from types less than 100 is on the TDB time scale while the time argument for types greater than 100 is on the TCB time scale.

\item Data types 901 through 910 have been reserved for the development of new ephemeris types by other groups.  Future versions of SPICE may recognize some of these types or other types developed by NAIF. Any decision to do so, and what identification number is assigned, is solely at the discretion of NAIF.

\item The ephemeris object numbers (\S\ref{ObjectIDNosApp}) $1\,000\,000\,001$, $1\,000\,000\,002$, and $1\,000\,000\,003$ have been reserved for Coordinate Time ephemerides. These ephemerides may store TT$-$TDB in the X-coordinate, TCG$-$TCB in the Y-coordinate, or both depending on the chosen object number (\S\ref{ObjectIDNos}).
\end{enumerate}

\section{Coordinate Time Ephemerides}\label{CoordTimeEphem}

The coordinate time scale for the ephemerides, either Barycentric Dynamical Time, TDB \citep{IAU2008}, or Barycentric Coordinate Time, TCB \citep{IAU2001a}, may be stored in a coordinate time ephemeris as Chebyshev polynomials. These time scales are stored either as TT~--~TDB or as TCG~--~TCB, where TT is Terrestrial Time \citep{IAU2001b} and TCG is Coordinate Geocentric Time \citep{IAU2001a}. A Type~2 SPK segment uses TDB and a Type~102 SPK segment uses TCB as the independent argument (\S\ref{Type2}). NAIF has assigned identification numbers (\S \ref{ObjectIDNos}) for coordinate time ephemerides:
\begin{itemize}
\item $1\,000\,000\,001 \Rightarrow$ TT$-$TDB data are stored in the $X$-coordinate
\item $1\,000\,000\,002 \Rightarrow$ TCG$-$TCB data are stored in the $Y$-coordinate
\item $1\,000\,000\,003 \Rightarrow$ TT$-$TDB data are stored in the $X$-coordinate and TCG$-$TCB data are stored in the $Y$-coordinate.
\end{itemize}
Data in the unused dimensions are set to 0 to prevent formatting errors that might occur when summary programs are used to display information about the contents of the SPK containing the segment. The second integer code, normally used to indicate the reference frame (\S\ref{RecognizedReferenceFrames}), is set to $1\,000\,000\,000$ to indicate the data being stored is a coordinate time ephemeris. 
 
The time bounds in the segment summary (\S\ref{SummaryRecords}) are {\em always} given on the TDB time scale. SPICE does not currently use or recognize the TCB time scale. CALCEPH does recognize and make use of kernels using the TCB time scale.

Binary PCKs, like SPKs, use NAIF identification numbers. But the Coordinate time ephemerides identification numbers are only recognized by SPK kernels.

\section{The Double Precision Array (DAF) Architecture}\label{DAF}

The SPK and binary PCK formats are based on the {\em Double precision Array File} (DAF) architecture. DAF was developed as a part of SPICE and written in ANSI standard FORTRAN~77\footnote{SPICE variants CSPICE, Icy, and Mice are also available and maintained by NAIF. These variants are designed to make SPICE available to programs written respectively in C, IDL, and Matlab. All three variants are derived from the original FORTRAN~77 code.}. Thus, some of the description of the DAF architecture and the SPK and binary PCK formats is derived directly from the FORTRAN~77 concepts of ``double precision'' and ``record length''.

A FORTRAN~77 {\tt DOUBLE PRECISION} data type is a floating point numerical value with a numerical precision of approximately 15 digits. It may also be designated {\tt REAL*8} in most dialects of FORTRAN~77. In Fortran~90 it is a {\tt real (kind = selected\_real\_kind(15))}, and is the C equivalent of a {\tt double}\footnote{Fortran~90 includes a real kind {\tt c\_long} to assure interoperability with C code. Whether or not {\tt c\_long} is identical to double precision is not specified.}.
 
Many Fortran files use a fixed record length. Storing and retrieving data from these file takes place in blocks with a constant, predetermined number of units. The unit size is platform dependent. Usually, it is in bytes. The DAF architecture includes a count of the units in the file that are called ``addresses''. A DAF address is {\em not} the same as a memory address, but is a method of locating data within a DAF architecture file (see \S\ref{ArrayAddr}). The number of records, double precision arrays, and number of elements an array can contain are limited by the number of words or bytes that can be addressed. Addresses are stored as 32-bit Fortran integers, which have a maximum positive value of 2,147,483,647.

The data in DAF architecture files are stored as ``arrays'' of {\tt DOUBLE PRECISION} numbers. DAF files are intended to be portable. Thus, the DAF architecture requires the array elements to consist of only {\tt DOUBLE PRECISION} numbers. These arrays may not contain equivalenced\footnote{The Fortran {\tt EQUIVALENCE} statement performs a function similar to a C {\tt union}. The FORTRAN~77 code
{\tt \begin{verse}
DOUBLE PRECISION X\\
INTEGER I(2)\\
EQUIVALENCE (X, I)\\
\end{verse}} 
\noindent could appear in C as
{\tt \begin{verse}
union data \{double x; int i{[2]};\} mydata;\\
\end{verse}} 
\noindent The stored order of the bytes of the equivalenced integers is dependent on the system architecture. The {\tt EQUIVALENCE} statement has been deprecated in Fortran~90.\label{Unionfootnote}} or encoded integer or character values.

\subsection{The Descriptive Summary and Segment Identifier}\label{DAFsumrec}

Each array contained in a DAF possesses a {\em descriptive summary}. These descriptive summaries are stored in the Summary Records (\S\ref{SummaryRecords}). The organization of a summary and its data are the same for each array in the DAF. The descriptive summary contains double precision and integer components. The number of double precision components, $ND$, and the number of integer components, $NI$, contained in the array summaries determine the array's format within the DAF architecture. The values for $ND$ and $NI$ are fixed when the array file is created. Any two array files that have the same values for $ND$ and $NI$ can be thought of as having the same ``format''\footnote{This does not mean that the individual arrays in a file contain the same kinds of data, only that they may be stored in the same DAF architecture file. Both the SPK and binary PCK specifications require that all of the arrays contain data pertinent to that kernel type.}. The values selected for $ND$ and $NI$ must satisfy
\begin{equation}\label{NDNIbounds}
2 \le NI \le 250 \;\;\; {\rm and} \;\;\; ND \le 125 - \frac{NI + 1}{2},
\end{equation}
where integer division is used, so $(NI + 1)/2$ is truncated. The final two integer components of the summary for an array are always the values of its initial and final addresses (\S\ref{ArrayAddr}), so $NI \geq 2$. How many values are needed and what they contain are left to the designer of a specific DAF type.

The double precision and integer values that describe each array are ``equivalenced'' into a double precision array before they are stored in the file\footnotemark[6]. The individual (unpacked) values are the {\em components} of the summary. The first $ND$ elements of the summary contain the double precision components. Each of the remaining elements contains a pair of integer components. If $NI$ is odd, the final element of the summary contains a single integer component.

Each array in an array file is further described by $NC$ characters of alphanumeric information called an {\em segment identifier}. These segment identifiers are stored in the Name Records (\S\ref{NameRecords}). The value of $NC$ is
\begin{eqnarray}
NC & = & 8 \left(ND + \frac{NI + 1}{2} \right)
\end{eqnarray}
using integer division. Most segment identifiers are short. For SPKs and binary PCKs they are 40 characters long. It is desirable, however, to make available alphanumeric data such as producer names, archive codes, historical data, or anything else not easily encoded as double precision or integer numbers. Thus, segment identifiers should not be used to replace comments in the Comment Area (\S\ref{CommentRecords}).

\subsubsection{Array Addresses}\label{ArrayAddr}

Every array file is a FORTRAN~77 direct access file, with a constant record length capable of storing up to 128 {\em words}. Each word consists of 64 bits and may contain one double precision number. The first record of a file contains words 1 through 128. The second record contains words 129 through 256, {\em etc.}. The number of each word is its {\em address} within the file.

The location of each array in an array file is defined by the {\em initial address} and {\em final address} of the array. These addresses are always the values of the final two integer components of the array's descriptive summary. This pair of addresses defines a contiguous set of words, which may fall within a single physical record or span a number of records. The elements of each array in an array file are stored in such a set. The initial address is the address of the first array element, and the array's final address is the address of the final array element.

The arrays in an array file form a doubly-linked list. A new array added to a file is placed at the tail of the list. The head and tail of the list can be located immediately. The other arrays can be located by moving a pointer through the list in either direction. The initial and final addresses may be used to access, retrieve, or update the entire array, or any contiguous set of elements therein.

\subsubsection{SPK and Binary PCK}

For both SPK kernels and binary PCK kernels the descriptive summary consists of  the two double precision and six integer values. The meaning of the values in the summary for a SPK kernel is discussed in \S\ref{segments}, and for a binary PCK kernel the meaning of these values are discussed in \S\ref{PCK}.

\subsection{Structure of an Array File}

A DAF array file is a Fortran direct access file. The record length is computer system dependent because different systems assign storage in different ways. For SPKs and PCKs the record length is set to contain 128 double precision numbers.

\subsubsection{Organization}\label{organization}

An array file contains five types of physical records:
\begin{enumerate}
\item A single {\em File Record} (\S\ref{TheFileRecord}). This record contains global information about the file.
\item A {\em Comment Area} (\S\ref{CommentRecords}) containing an arbitrary number of {\em Comment Records}. These records allow the user to store information about the data within the array file. Typical information might include the source of the data, or the names of programs used to process and interpret it.
\item {\em Summary Records} (\S\ref{SummaryRecords}). These contain array descriptive summaries (\S\ref{DAFsumrec}) and pointers to other Summary Records. The number of Summary Records in a particular array file is a function of the number of arrays stored in the file.
\item {\em Name Records} (\S\ref{NameRecords}). These contain a character string identifier for each array. An array file contains one Name Record for each Summary Record. The length of a name is determined by the number of values stored in the Summary record as described in \S\ref{DAFsumrec}.
\item An arbitrary number of {\em Element Records} (\S\ref{ElementRecords}). These contain the array data stored in the array file.
\end{enumerate}

\subsubsection{The File Record}\label{TheFileRecord}

The File Record is always the first physical record in an array file. It contains nine items in the following order:
\begin{enumerate}
\item An {\em identification word}, ``\texttt{yyy/xxxx}'', where ``\texttt{yyy}'' is a three character string indicating the file architecture and ``\texttt{xxxx}'' is a four character string indicating the type of data stored in the array file. For an SPK the file identification word is ``\texttt{DAF/SPK }'', for a text PCK it is ``\texttt{KPL/PCK }'', and for a binary PCK it is ``\texttt{DAF/PCK }''.
\item The value of $ND$, the number of double precision components in each summary (\S\ref{DAFsumrec}).
\item The value of $NI$, the number of integer components in each summary (\S\ref{DAFsumrec}).
\item A 60 character internal name for the array file.
\item The record number of the initial Summary Record in the file.
\item The record number of the final Summary Record in the file.
\item The first free address in the file: the address where the first element of the next array added to the file will be stored.
\item A binary file format identification string.
\item An FTP transmission corruption test string.
\end{enumerate}
The numerical values in the File Record are all stored as binary integers. The portion of the File Record which does not contain data is padded with the ASCII null character, $<000>$\footnote{Throughout \S\ref{TheFileRecord}, the value between the delimiters `$<$' and `$>$' is the numeric value of that byte.}.

\paragraph{The Binary File Format Identification String:}

The binary file format identification string is an eight character string identifying the order in which the binary data bytes were stored when written and the interpretation of the bits within the floating point number. Currently, there are four recognized formats:
\begin{enumerate}
\item ``$BIG-IEEE$'': The IEEE format for floating point mantissa and exponent stored in big-endian order from most significant to least significant.
\item ``$LTL-IEEE$'': The IEEE format for floating point mantissa and exponent stored in little-endian order from least significant to most significant.
\item ``$VAX-GFLT$'': Integers are stored in little-endian order, single precision floating  point numbers use the VAX F format, and double precision numbers use the VAX G format \citep{NSSDC2014a}.
\item ``$VAX-DFLT$'': Integers are stored in little-endian order, single precision floating  point numbers use the VAX F format, and double precision numbers use the VAX D format \citep{NSSDC2014a}.
\end{enumerate}
Other formats are available \citep{NSSDC2014b}, but these four formats predominate. In particular, little-endian order is the most common due to the pervasive use of the Intel x86 architecture machines, which use a little-endian architecture \citep{Intel2004}.

\paragraph{The FTP Transmission Corruption Test String:}

If the user neglects to invoke the IMAGE (BINARY) transfer mode when transferring a binary file from one platform to another using the FTP protocol, an ASCII mode transfer may occur, and the file may become corrupted. The most likely corruption of a binary file is the possible substitution of one set of line terminators for another. Placing a string that is a representative set of character sequences that are susceptible to corruption in the File Record makes it possible to trap and report any problems to the user if corrupted kernels are loaded at run time. Moving test binary files from one platform to another shows that the clusters of ASCII codes most likely to be corrupted are:
\begin{itemize}
\item $<013>$ -- The text line terminator on older Macintosh-based platforms.
\item $<010>$ -- The text line terminator on UNIX-based platforms.
\item $<013><010>$ -- The text line terminator on Microsoft platforms.
\item $<013><000>$ -- This sequence of characters maps into $<013>$ on some UNIX-based systems (HP, SGI, NEXT).
\item $<129>$ -- Macintosh based systems permute ASCII values whose parity bit is set. Thus, ASCII values greater than 127 are altered.
\item $<016><206>$ -- Some older FTP servers running Microsoft operating systems convert this sequence of ASCII codes to $<016><016><206>$.
\end{itemize}
These clusters may not be the complete set of clusters that may be corrupted through an improper FTP transfer.

The substitution of one set of line terminators for another may result in expansion or compression of certain sequences of bytes. If the clusters are juxtaposed, new sequences of adjacent bytes, also subject to transformation, might be formed. The FTP transmission corruption test string is inserted so that it can be located in the event compression or expansion, either within the test string itself, or elsewhere in the file record, shifts it away from its default location. It also must include a mechanism to prevent interaction between the clusters. The solution is to bracket the entire test string with the start and stop identifiers `FTPSTR' and `ENDFTP', and separate the clusters with the printable delimiter `:'. The FTP transmission corruption test string, inserted into the File Record starting with the 700th character, is:
\begin{verse}
\item FTPSTR:$<013>:<010>:<013><010>:$\\
$<013><000>:<129>:<016><206>$:ENDFTP
\end{verse}
This string may be modified in the future if other clusters of ASCII codes likely to be corrupted by an improper FTP transfer are discovered.

\subsubsection{The Comment Area}\label{CommentRecords}

The contents and formats of Comment Area are left entirely to the user, and may be left empty. The initial Comment Record is the second record of the file, and the final Comment Record immediately precedes the initial Summary Record of the file. Comment Records may be used to store any data desired by the file constructor. The Comment Area may contain {\em only} printable ASCII characters, specifically ASCII 32-126.
 
The only limit on the line length in the Comment Area is: The number of characters must be representable by a FORTRAN integer. SPICE supports a line length of up to 255 characters. A shorter maximum line length, however, may enhance readability.

\subsubsection {Summary Records}\label{SummaryRecords}

A Summary Record contains a maximum of 128 double precision words. The first three words of each Summary Record are reserved for the {\em control items}. They are:
\begin{enumerate}
\item The record number of the next Summary Record in the file. (If this is the final Summary Record then the value is 0.)
\item The record number of the previous Summary Record in the file. (If this is the initial Summary Record then the value is 0.)
\item The number of summaries stored in this record.
\end{enumerate}
Although the control items are integer values, they are stored as double precision numbers. This allows Summary Records to be buffered using the same mechanism as Element Records, which contain only double precision numbers.

\paragraph{The Summaries:}

The control items are followed immediately by the summaries. The number of summaries, $NS$, that can fit in a single Summary Record depends on the summary size $SS$, which is a function of $NI$ and $ND$ (\S\ref{TheFileRecord}). Using integer division,
\begin{equation}
SS  =  ND + \frac{NI + 1}{2}.
\end{equation}
Then
\begin{equation}
NS = \frac{125}{SS}.
\end{equation}

\begin{figure}[h]
\centering
\includegraphics[width = \textwidth]{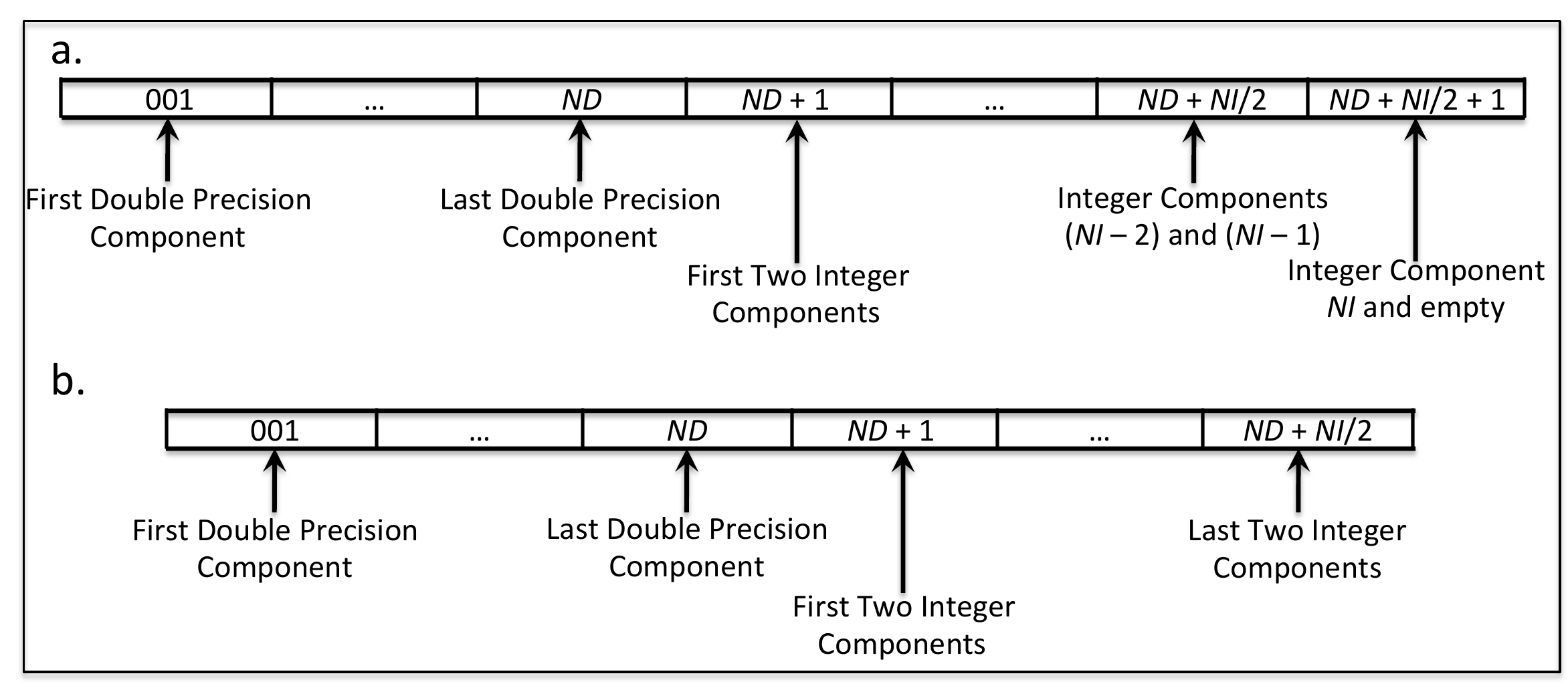}
\caption{{\bf a}. An example of the format of words in a summary with an odd number of integer components. {\bf b}. An example of the format of words in a summary with an even number of integer components.}\label{SPKFig0}
\end{figure}

Each summary contains the $ND$ double precision components followed by the $NI$ integer components of the summary. The integer components are stored, in pairs, as equivalenced {\tt DOUBLE PRECISION} numbers (\S\ref{DAFsumrec}). Thus, the integer components require a total of $(NI + 1) / 2$ words of space, and the byte order depends on the system architecture. Figure~\ref{SPKFig0}a shows the layout of the data within a summary with an odd number of integer components, and Fig.~\ref{SPKFig0}b shows the layout of the data within a summary with an even number of integer components.

The values of most of these components are up to the designer of the DAF type. The final two integer components, however, contain the initial and final addresses of the array (\S\ref{ArrayAddr}).

Summaries are never split across physical record boundaries. Thus, if the number of remaining bytes in the Summary Record are insufficient to hold a summary, they remain unused.

\begin{figure}[h]
\centering
\includegraphics[width = \textwidth]{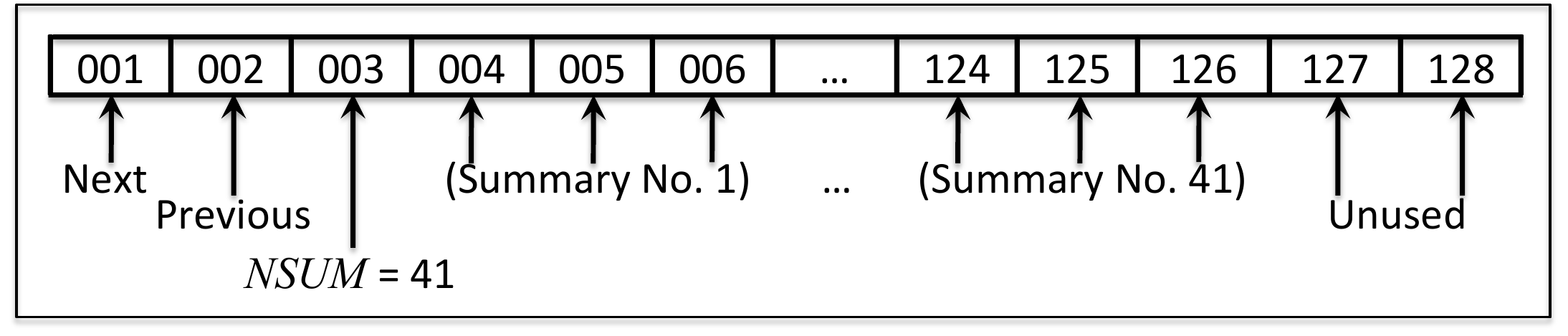}
\caption{An example of the format of words in a Summary Record. The record is divided into double precision words and, in this case, the number of words in a summary, $SS$, is three.} \label{SPKFig1}
\end{figure}

An example of a Summary Record is depicted in Fig.~\ref{SPKFig1}. The value in each box corresponds to the address of the double precision word in the record. Let $SS = 3$, so
\begin{eqnarray}
NS = \frac{125}{3} = 41\nonumber
\end{eqnarray}
summaries can fit in the Summary Record. The first summary is stored in words 4 through $SS + 3$, the second summary is stored in words $SS + 4$ through $2 SS + 3$, and so on. This uses a total of 126 words, leaving two words empty.

\subsubsection{Name Records}\label{NameRecords}

Each Name Record contains a set of character strings to identify the arrays. A Name Record always follows a Summary Record so a new Name Record is added to the kernel each time a new Summary Record is added. The number of names in a Name Record is equal to the number of summaries in the corresponding Summary Record. The maximum length for a name in the Name Record, $NC$ is
\begin{equation}\label{NCEq}
NC = 8 \left(ND + \frac{NI + 1}{2} \right) = 8\; SS.
\end{equation}

\begin{figure}[h]
\centering
\includegraphics[width = \textwidth]{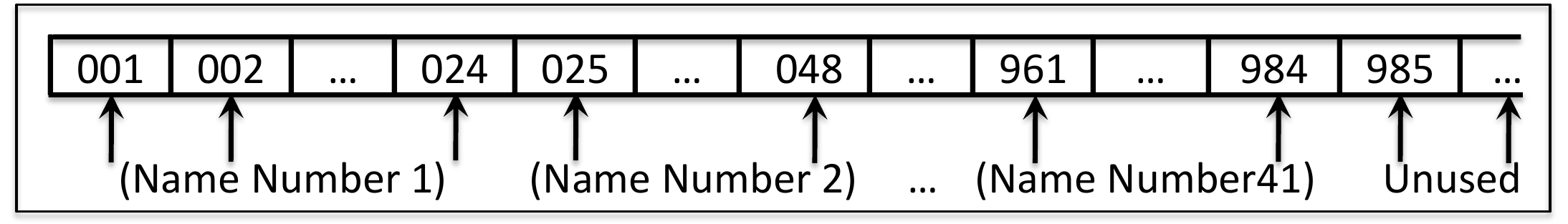}
\caption{An example of the format of characters in a Name Record. The record is divided into characters and, in this case, the number of characters in a name, $NC$, is $8\; SS = 24$.}\label{SPKFig2}
\end{figure}

An example of a Name Record is depicted in Fig.~\ref{SPKFig2}. The numbers correspond to the number of the characters in the record. The first name is stored in characters 1 through $NC$; the second name is stored in characters $NC + 1$ through $2\; NC$; and so on. In this example $SS = 3$, so $NS = 41$ and
\begin{eqnarray}
 NC = 8 \times 3 = 24\nonumber
 \end{eqnarray}
characters can fit in a name. This uses a total of 984 characters, leaving the characters beginning with 985 empty.

\subsubsection{Element Records}\label{ElementRecords}

Most of the records in an array file are Element Records. Element Records hold the elements of the arrays stored in the file. Each Element Record has room for 128 double precision numbers. A record that immediately precedes a Summary Record or is the last record of a file may be partially filled.

All elements belonging to the same array are stored contiguously. An array may span multiple Element Records.  If an array extends beyond the end of an Element Record, the element immediately following the last address in that Element Record is placed in the first address of the next Element Record.

\begin{figure}[h]
\centering
\includegraphics[width = \textwidth]{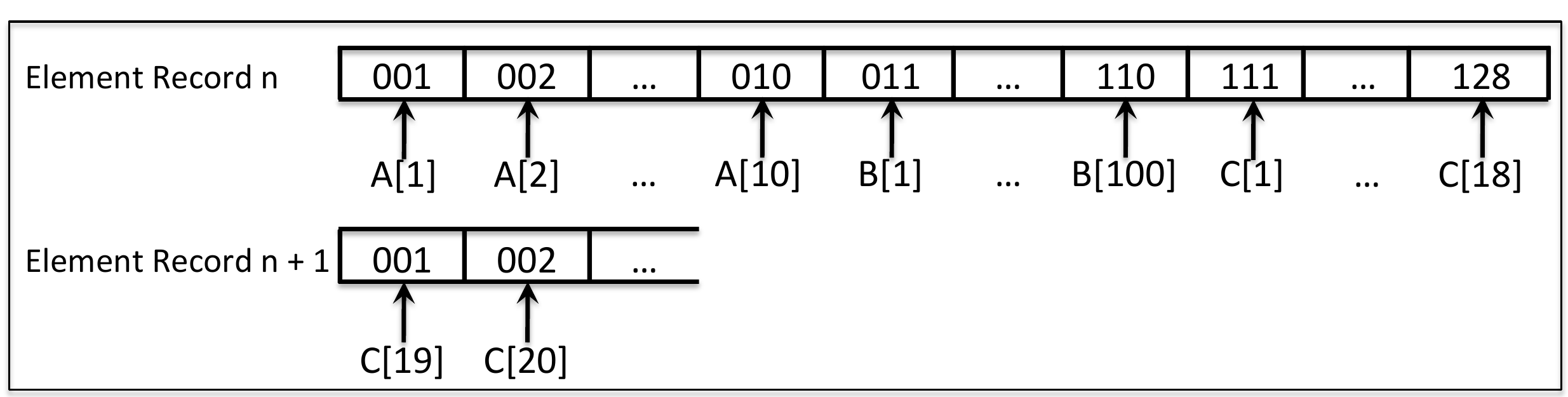}
\caption{An example of an Element Record. The record is divided into double precision words. In this case, the record contains three arrays, $A$, $B$, and $C$. Array $A$ contains 10 elements, array $B$ 100 elements, and array $C$ 20 elements.}\label{SPKFig3}
\end{figure}

The elements stored in an Element Record may belong to more than one array. Fig.~\ref{SPKFig3} shows an Element Record containing three arrays: $A$, $B$, and $C$. Array $A$ has 10 elements, array $B$ has 100 elements, and array $C$ has 20 elements.

\subsubsection{An SPK Example}

This example illustrates the use of addresses and lists within an SPK showing how one file may be created, and how arrays may be added to that file.

The following notation will be used for items defined in the File Record:
\begin{itemize}
\item {\em IDWORD} is the identification word,
\item $ND$ and $NI$ are the number of double precision and integer values used to define the format of the file,
\item $RI$ and $RF$ are the record numbers of the initial and final Summary Records in the file, and
\item $FFA$ is the first free address in the file.
\item $NEXT$ and $PREV$ are the record numbers of the next and previous Summary Records in the file, and
\item $NSUM$ is the number of summaries stored in the record.
\end{itemize}
Details on these parameters may be found in \S\ref{TheFileRecord} and \S\ref{SummaryRecords}.

The {\em IDWORD} written to the new file is the concatenation of the string ``DAF/'' with the kernel type string, and must contain eight characters. For an SPK, the {\em IDWORD} written to the new file is ``DAF/SPK\hspace{6pt}'' where the eighth character of the {\em IDWORD} is a {\em blank} character.

For SPK kernels, the value for $ND = 2$ and for $NI = 6$. The meanings for the double precision and integer components of an SPK Summary Record are discussed in \S\ref{segments}. The summary size, $SS$ is
\begin{eqnarray}
SS = ND + \frac{NI + 1}{2} = 2 + 3 = 5,\nonumber
\end{eqnarray}
so each segment summary requires 5 double precision words of storage, so $NSUM_{Max}$, the number of summaries each Summary Record can hold
\begin{eqnarray}
NSUM_{Max} = \frac{125}{SS} = \frac{125}{5} = 25\nonumber
\end{eqnarray}
summaries.

\begin{figure}[h]
\centering
\includegraphics[width = \textwidth]{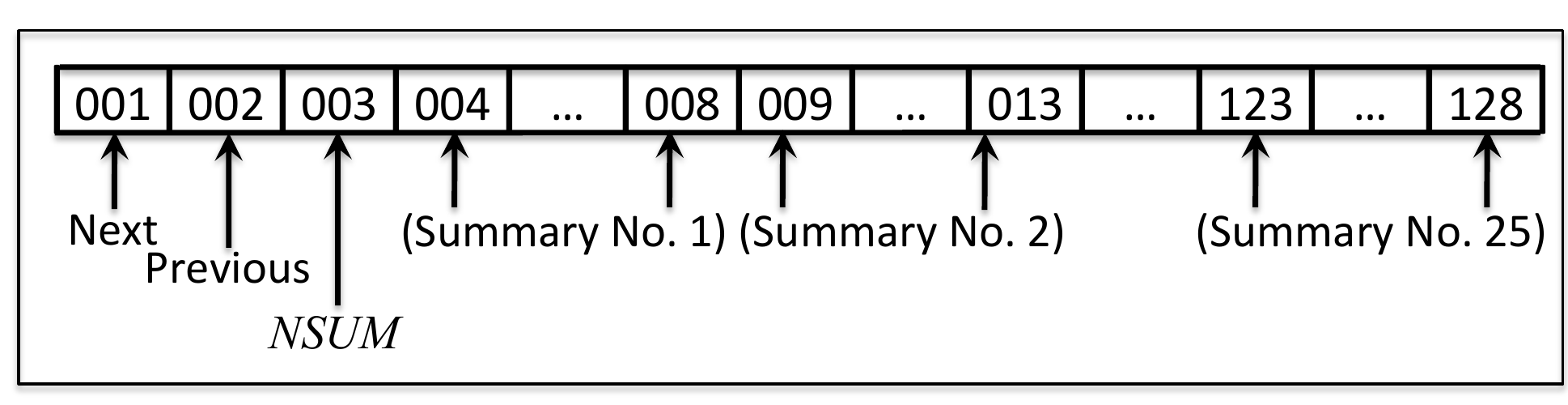}
\caption{An example of an SPK Summary Record where $ND = 2$ and $NI = 6$, so $SS = 5$ and $NSUM_{Max} = 25$.} \label{SPKFig4}
\end{figure}

Figure~\ref{SPKFig4} shows the layout of a typical SPK Summary Record.

\begin{figure}[h]
\centering
\includegraphics[width = \textwidth]{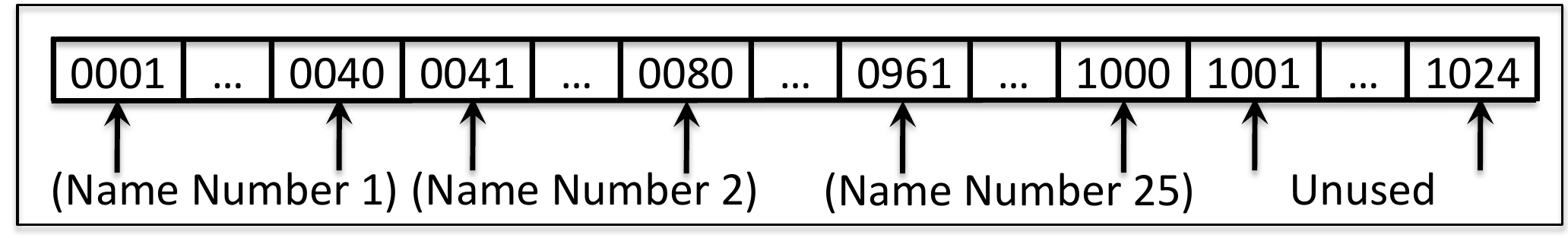}
\caption{An example of an SPK Name Record where $ND = 2$ and $NI = 6$. Thus, $NSUM_{Max} = 25$ so $NC = 40$.} \label{SPKFig5}
\end{figure}

The number of names that a Name Record can hold is equal to the number of summaries that the Summary Record can hold. Here it is 25, and the number of characters in each name array is
\begin{eqnarray}
NC = 8\, SS = 8 \times 5 = 40.\nonumber
\end{eqnarray}
Thus, the Name Record, which immediately follows each Summary Record, has space reserved as shown in Fig.~\ref{SPKFig5}.

Assume that the number of Comment Records is 10. Then, the File Record information is stored in record 1, the Comment Record data are stored in records 2 through 11, and the initial Summary Record is record 12. If the file is empty, the initial Summary Record, $RI$, is also the final Summary Record, $RF$:
\begin{eqnarray}
RI = RF = 12\nonumber
\end{eqnarray}
The first Name Record for the file is immediately after the Summary Record, in record 13. Thus, the first free address, $FFA$, in the file is the first word in record 14:
\begin{eqnarray}
FFA  =  w + (r - 1) * 128 =  1 + (14 - 1) * 128 =  1665\nonumber
\end{eqnarray}
where $w$ is the word number and $r$ is the record number.

The File Record also contains an internal file name of up to 60 characters. For this example the internal file name will be ``TESTFILE''. For the rest of this example, the binary file format identification string and the FTP transmission corruption string (\S\ref{TheFileRecord}) will be ignored. The File Record will be represented by a collection of values enclosed by braces and preceded by its record number, $r$:
\begin{verse}
$r$ \{ $IDWORD = x$, $ND = a$, $NI = b$, $IFNAME = c$, $RI = d$,
$RF = e$, $FFA = f$  \}.
\end{verse}
where $x$, $a$, $b$, $c$, $d$, $e$, and $f$ are appropriate values. The File Record for this example file is initially:
\begin{verse}
1 \{ $IDWORD =$ ``DAF/SPK\hspace{6pt}'', $ND = 2$, $NI = 6$, $IFNAME =$
``TESTFILE'', $RI = 12$, $RF = 12$, $FFA = 1665$ \}.
\end{verse}

There is only one Summary Record, and the file contains no arrays. Thus, the values of $NEXT$, $PREV$, and $NSUM$ in the Summary Record are all initially zero. For the rest of the example, each Summary Record will be represented by a collection of values enclosed by angle brackets and preceded by its record number:
\begin{verse}
$r$ $<$ $NEXT = a$, $PREV = b$, $NSUM = c$, ($d$, $e$),
($f$, $g$), ..., ($h$, $i$) $>$
\end{verse}
where $a$, $b$, and $c$ are integers and the ordered pairs enclosed in parentheses are the initial and final addresses of the arrays whose summaries are contained in the record. The remaining components of each summary are ignored to make the example easier to follow. Thus, the first, and so far only, Summary Record for this example file is initially:
\begin{verse}
12 $<$ $NEXT = 0$, $PREV = 0$, $NSUM = 0$, (0, 0), (0, 0), ..., (0, 0) $>$.
\end{verse}

Name Records will be represented by
\begin{verse}
$r$ $<$ ``$n$'' $>$,
\end{verse}
where $n$ is the number of names it currently contains, and Element Records will be represented by
\begin{verse}
$r$ $< N >$
\end{verse}
where $N$ is the number of elements stored in the record.

Once the initial Summary and Name Records have been written, the file looks like this:
\begin{verse}
\hspace{1 mm} 1 \{ $IDWORD =$ ``DAF/SPK\hspace{6pt}'', $ND = 2$, $NI = 6$, $IFNAME =$ ``TESTFILE'', $RI = 12$, $RF = 12$, $FFA = 1665$ \}\\
\hspace{1 mm} 2\\
.\\
. {\em Records 2 through 11 are the comment area.}\\
.\\
11\\
12 $<$ $NEXT = 0$, $PREV = 0$, $NSUM = 0$, (0, 0), (0, 0), ..., (0, 0)$ >$\\
13 $<$ ``0'' $>$
\end{verse}

An array $A_1$, containing 100 elements, is added to the file. This array will be stored contiguously, beginning at the first free address. Thus, its initial and final addresses will be 1665 and 1764, respectively. The entire array fits into a single record, so one Element Record is added to the file. The value of $NSUM$ in the Summary Record is incremented by one. The new value of $FFA$ is the address following the final address of the new array: 1765. After $A_1$ is added, the file is now
\begin{verse}
\hspace{1 mm} 1 \{ $IDWORD =$ ``DAF/SPK\hspace{6pt}'', $ND = 2$, $NI = 6$, $IFNAME =$ ``TESTFILE'', $RI = 12$, $RF = 12$, $FFA = 1765$ \}\\
\hspace{1 mm} 2\\
.\\
. {\em Records 2 through 11 are the comment area.}\\
.\\
11\\
12 $<$ $NEXT = 0$, $PREV = 0$, $NSUM = 1$, (1665, 1764), (0, 0),
..., (0, 0) $>$\\
13 $<$ ``1'' $>$\\
14 $<$ 100 $>$ {\em 100 words for $A_1$.}
\end{verse}

A second array $A_2$, containing 200 elements, is added to the file. The initial and final addresses of $A_2$ will be 1765 and 1964. This array will fill the remainder of the first Element Record, all of the second record, and part of the third. Thus, two Element Records are added to the file. The value of $NSUM$ in the Summary Record is again incremented, and the new value of $FFA =1965$ is stored in the File Record. The file is now:
\begin{verse}
\hspace{1 mm} 1 \{ $IDWORD =$ ``DAF/SPK\hspace{6pt}'', $ND = 2$, $NI = 6$, $IFNAME =$ ``TESTFILE'', $RI = 12$, $RF = 12$, $FFA = 1965$ \}\\
\hspace{1 mm} 2\\
.\\
. {\em Records 2 through 11 are the comment area.}\\
.\\
11\\
12 $<$ $NEXT = 0$, $PREV = 0$, $NSUM = 2$, (1665, 1764), (1765, 1964),
..., (0, 0) $>$\\
13 $<$ ``2'' $>$\\
14 $<$ 128 $>$ {\em 100 words for $A_1$, 28 words for $A_2$}\\
15 $<$ 128 $>$ {\em 128 words for $A_2$}\\
16 \hspace{1 mm} $<$  44 $>$ \hspace{1 mm} {\em 44 words for $A_2$}
\end{verse}

For arrays $A_3$ through $A_{24}$, each containing 10 elements, the process is repeated. These will take up addresses 1965 through 2184. They will fill the remainder of the third Element Record, all of the fourth Element Record, and the first 48 words of the fifth Element Record. The file is now:
\begin{verse}
\hspace{1 mm} 1 \{ $IDWORD =$ ``DAF/SPK\hspace{6pt}'', $ND = 2$, $NI = 6$, $IFNAME =$ ``TESTFILE'', $RI = 12$, $RF = 12$, $FFA = 2185$ \}\\
\hspace{1 mm} 2\\
.\\
. {\em Records 2 through 11 are the comment area.}\\
.\\
11\\
12 $<$ $NEXT = 0$, $PREV = 0$, $NSUM = 24$, (1665, 1764), (1765,
1964), ..., (2175, 2184), (0, 0) $>$\\
13 $<$ ``24'' $>$\\
14 $<$ 128 $>$ {\em 100 words for $A_1$, 28 words for $A_2$}\\
15 $<$ 128 $>$ {\em 128 words for $A_2$}\\
16 $<$ 128 $>$ \hspace{1 mm} {\em 44 words for $A_2$, \hspace{1 mm} 10 words each for $A_3$--$A_{10}$, \hspace{1 pt} and 4 words of $A_{11}$}\\
17 $<$ 128 $>$ \hspace{3 mm} {\em 6 words for $A_{11}$, 10 words each for $A_{12}$--$A_{23}$, and 2 words of $A_{24}$}\\
18 \hspace{1 mm} $<$  48 $>$ \hspace{3 mm} {\em 8 words for $A_{24}$.}
\end{verse}

Let array $A_{25}$ contain 150 elements. Its initial and final addresses are 2185 and 2334. The array fills the remainder of the fifth Element Record, and part of a sixth, so one new Element Record is added. And the value of $NSUM$ is in the Summary Record is incremented. The final Summary Record is now full. Thus, the following adjustments are made to allow new data to be added to the file:
\begin{enumerate}
\item New Summary and Name Records are added to the file.
\item The value of $NEXT$ in the old Summary Record is set to the record number of the new Summary record, in this case 20.
\item The value of $PREV$ in the new Summary Record is set to the record number of the old Summary record, in this case 12.
\item The File Record is updated so that the value of $RF$ points to the new final Summary Record.
\item The value of $FFA$ in the File Record points to address 2689, the first word in the first Element Record following the new Name Record.
\end{enumerate}
The previous Element Record, record 19, will remain only partially filled.
\begin{verse}
\hspace{1 mm} 1 \{ $IDWORD =$ ``DAF/SPK\hspace{6pt}'', $ND = 2$, $NI = 6$, $IFNAME =$ ``TESTFILE'', $RI = 12$, $RF = 20$, $FFA = 2689$ \}\\
\hspace{1 mm} 2\\
.\\
. {\em Records 2 through 11 are the comment area.}\\
.\\
11\\
12 $<$ $NEXT {\scriptstyle =} 20$, $PREV {\scriptstyle =} 0$, $NSUM {\scriptstyle =} 3$, (1665, 1764), (1765, 1964), ..., (2185, 2334)$>$\\
13 $<$ ``25'' $>$\\
14 $<$ 128 $>$ {\em 100 words for $A_1$, 28 words for $A_2$}\\
15 $<$ 128 $>$ {\em 128 words for $A_2$}\\
16 $<$ 128 $>$ \hspace{1 mm} {\em 44 words for $A_2$, \hspace{1 mm} 10 words each for $A_3$--$A_{10}$, \hspace{1 pt} and 4 words of $A_{11}$}\\
17 $<$ 128 $>$ \hspace{3 mm} {\em 6 words for $A_{11}$, 10 words each for $A_{12}$--$A_{23}$, and 2 words of $A_{24}$}\\
18 $<$ 128 $>$ \hspace{3 mm} {\em 8 words for $A_{24}$ \hspace{1 pt} and 120 words for $A_{25}$}\\
19 \hspace{1 mm} $<$  30 $>$ \hspace{1 mm} {\em 30 words for $A_{25}$}.\\
20 $<$ $NEXT = 0$, $PREV = 12$, $NSUM = 0$, (0, 0), (0, 0), ..., (0, 0) $>$\\
21 $<$ ``0'' $>$
\end{verse}

Adding more arrays then continues with a new Element Record as record 22. As additional data are added to the file:
\begin{itemize}
\item Element Records are added as necessary.
\item Summary and Name Records are updated.
\item When the final Summary Record is filled:
   \begin{itemize}
   \item New Summary and Name Records are added.
   \item The value of $RF$ is updated.
   \end{itemize}
\item The value of $FFA$ is updated.
\end{itemize}

\section{The SPK Format}\label{SPK}

The SPK format was developed specifically to store position ephemerides of objects. Each SPK kernel contains a number of segments (\S\ref{segments}) consisting of an ephemeris for a single object. The file identification word for an SPK kernel is ``\texttt{DAF/SPK }''.

\subsection{Segments}\label{segments}

An SPK {\em segment} consists of a DAF array in an SPK kernel. A kernel may contain one or more segments. Each segment contains data sufficient to determine the ephemeris of an object in a specified reference frame. All the data in a segment consists of a single SPK type, but multiple SPK types may occur in a single SPK kernel. The associated segment summary (\S\ref{DAFsumrec}) has two double precision components ($ND = 2$), and six integer components ($NI = 6$). Thus, the maximum number of characters in a name is (using integer arithmetic)
\begin{eqnarray}
NC & = & 8 \left(2 + \frac{6 + 1}{2} \right) = 40\nonumber
\end{eqnarray}
characters.

The double precision components of the summary are:
\begin{enumerate}
\item the initial epoch and
\item the final epoch
\end{enumerate}
of the interval for which data are contained in the segment, on the TDB time scale from J2000.0 (JD~245\-1545.0). The initial and final epochs are {\em always} given as seconds from J2000.0 even when the time argument used in the segment is TCB. The time argument for the data in the segment, when expressed using the TDB time scale, must be between these two values.

The integer components of the summary are:
\begin{enumerate}
\item the NAIF object identification number (\S\ref{ObjectIDNos}) of the target    body,
\item the NAIF object identification code of the center,
\item the NAIF integer code (\S\ref{RecognizedReferenceFrames}) for the    reference frame,.
\item the integer code for the SPK data type (\S\ref{SPKTypes}),
\item the initial address of the array, and
\item the final address of the array.
\end{enumerate}
The first two integer components of the summary are discussed in \S\ref{ObjectIDNos} The third integer component is discussed in \S\ref{RecognizedReferenceFrames}. The fourth integer component is discussed in \S\ref{SPKTypes}. And the last last two integer components are discussed in \S\ref{ArrayAddr}.

Segments within an SPK need not be ordered according to time. Segments providing data for a later time period may precede segments covering an earlier time period. In SPICE, segment order implies priority. For a given object, segment priority increases with distance from the beginning of the file: segments closer to the end of the file have higher priority than segments for the same target body that occur earlier in the file. When a data request for a specified target body is made, the segment whose time interval includes the requested time with highest priority will be selected to satisfy the request. This priority scheme will cause a higher priority segment for a target body to mask a lower priority segment for the same body over the intersection of the coverage intervals of the two segments.

\subsubsection{NAIF Object Identification Numbers}\label{ObjectIDNos}

The first integer component of the summary is the NAIF object identification number for the ephem\-eris object. Each object in the solar system is assigned a unique identification  number. For example, the identification number for Jupiter is 599 and the identification number for the Sun's center is 10. Thus, the target body and center for the heliocentric ephemeris of Jupiter are 599 and 10, respectively. Similarly, the identification number for Jupiter's barycenter\footnote{There is a conceptual difference between the center of mass of a planet and a planet satellite system even when the planet has no satellites (\S\ref{Barycenters}) Hence, the ephemeris of a planet is {\em not} identical to the ephemeris of a planet-satellite system barycenter and require separate object identification numbers.} is 5 and the identification number for the solar system barycenter is 0. Thus, the target body and center for the heliocentric ephemeris of Jupiter's barycenter about the solar system barycenter are 5 and 0, respectively. Appendix~\ref{ObjectIDNosApp} details how these numbers are assigned. The tables in Appendix~\ref{ObjectIDNosApp} give the available identification numbers recognized by SPICE at the time this report was written.
 
Often it is more convenient to give the ephemeris of an object with respect to a nearby object rather than with respect to the origin of the reference frame. For example, the Moon's ephemeris may be referred to the ICRS but the coordinates may be given with respect to the geocenter. The second integer component is the NAIF identification number for the reference object from which the ephemeris object's ephemeris is offset.

\subsubsection{Recognized Reference Frames}\label{RecognizedReferenceFrames}

The third integer component of the SPK summary is the NAIF code for the reference frame. Reference frames may be either inertial or non-inertial.

Most of the inertial reference frames recognized are for specific JPL ephemerides and are of little interest in the present context. The identification number for the International Celestial Reference System, henceforth ICRS\footnote{Although it is called the International Celestial Reference {\em System}, the ICRS is consistent with the object called a reference {\em frame} in SPICE.}, is 1 and its name is ``J2000''. Other currently recognized, ``built-in'' inertial reference frame names and their identity numbers, found in \citet{Semenov2010b}, are given in Table~\ref{InertialFramesTable}.

\begin{table}
\begin{center}
\caption{Selected available inertial reference frame names \citep{Semenov2010b}.\label{InertialFramesTable}}
\begin{tabular}{l l r}
\hline\hline
\multicolumn{1}{c}{Frame Name} & \multicolumn{1}{c}{Description} & \multicolumn{1}{c}{ID No.} \\
\hline
J2000      & Earth mean equator, dynamical equinox of J2000.0$^{a}$ & 1 \\
B1950      & Earth mean equator, dynamical equinox of B1950.0 & 2 \\
FK4        & Fundamental Catalog (4) & 3 \\
GALACTIC   & Galactic System II & 13 \\
ECLIPJ2000 & Earth mean ecliptic and equinox of the Julian epoch J2000.0 & 17 \\
ECLIPB1950 & Earth mean ecliptic and equinox of the Besselian epoch B1950.0 & 18 \\
\hline
\multicolumn{3}{l}{$^{a}${\footnotesize This reference frame is taken to be identical with the International Celestial Reference}}\\
\multicolumn{3}{l}{{\footnotesize System (ICRS).}}
\end{tabular}
\end{center}
\end{table}

The currently recognized ``built-in'' non-inertial reference frame names are found in \citet{Semenov2010b}. The identity numbers of those reference frames of interest are tabulated in Table~\ref{NonInertialFramesTable}.
 
\begin{table}
\begin{center}
\caption{Selected available non-inertial reference frame names \citep{Bachman2007,Bachman2008}.}\label{NonInertialFramesTable}
\begin{tabular}{l l r}
\hline\hline
\multicolumn{1}{c}{Frame Name} & \multicolumn{1}{c}{Description} & \multicolumn{1}{c}{ID No.} \\
\hline
IAU\_EARTH      & The geocenter    & 10013\\
IAU\_MARS       & Mars' center     & 10014\\
IAU\_JUPITER    & Jupiter's center & 10015\\
IAU\_SATURN     & Saturn's center  & 10016\\
IAU\_URANUS     & Uranus' center   & 10017\\
IAU\_NEPTUNE    & Neptune's center & 10018\\
IAU\_PLUTO      & Pluto's center   & 10019\\
ITRF93          & International Terrestrial Reference Frame 1993$^{a}$ & 13000 \\
MOON\_PA        & Generic lunar axes of the principal moments of inertia$^{b}$ & 31000 \\
MOON\_ME        & Generic lunar mean Earth-mean pole of rotation axes$^{b}$ & 31001 \\
MOON\_PA\_DE421 & DE421 lunar axes of the principal moments of inertia & 31006 \\
MOON\_ME\_DE421 & DE421 lunar mean Earth-mean pole of rotation$^{c}$ & 31007 \\
\hline
\multicolumn{3}{l}{$^{a}${\footnotesize Available as a generic kernel from NAIF.}}\\
\multicolumn{3}{l}{$^{b}${\footnotesize The generic lunar orientation ephemerides are not associated with a specific lunar ephemeris. They}}\\[-3pt]
\multicolumn{3}{l}{\hspace{6pt}{\footnotesize may be linked to a specific ephemerides by SPICE, \S\ref{BinaryPCKFrames}\citep{Bachman2008}.}}\\
\multicolumn{3}{l}{$^{c}${\footnotesize A constant offset rotation matrix from the axes of the lunar principal moments of inertia is used to}}\\[-3pt]
\multicolumn{3}{l}{{\hspace{6pt}\footnotesize  specify the lunar mean Earth-mean pole of rotation reference frame \citep{Bachman2008}.}}\\
\end{tabular}
\end{center}
\end{table}

\citet{Semenov2010b} also describes how to specify additional reference frames. These reference frames may be either inertial or non-inertial.

\subsection{SPK Chebyshev Data Types}\label{SPKTypes}

SPK kernels use multiple data types to store the ephemeris data. Like the solar system bodies, each data type is recognized by an identification number. SPICE currently has reserved identification numbers for 24 data types in the SPK. These are types 1 through 21, 102, 103, and 120. Kernels that do not use the correct format for these data types will not be correctly interpreted by SPICE or other readers of these kernels. NAIF may also reserve other data types in the future. SPK data type identifiers 901 through 910 have been set aside for user defined data types.

Only the six types that use Chebyshev polynomials of the first kind \citep[e.g.][]{Rivlin1974} are of interest here\footnote{These types are also used for binary PCKs; however, they are used to store orientation data in binary PCK (\S\ref{BinaryPCK}) and positional data in SPKs.}. They are:
\begin{itemize}
\item Type 2. Chebyshev polynomials (position only): These are sets of coefficients for the {\em X}, {\em Y}, and {\em Z} rectangular coordinates of the body position. The velocity of the body is obtained by differentiation of the Chebyshev polynomials. The time argument for these ephemerides uses the TDB time scale. The details of ordering of data in Type~2 kernels is given in \S\ref{Type2}.
\item Type 3. Chebyshev polynomials (position and velocity): These are sets of coefficients for the {\em X}, {\em Y}, and {\em Z} rectangular coordinates of the body position, and the corresponding components of the velocity. The time argument for these ephemerides uses the TDB time scale. The details of ordering of data in Type~3 kernels is given in \S\ref{Type3}.
\item Type 20. Chebyshev polynomials (velocity only): These are sets of coefficients for the $\dot{X}$, $\dot{Y}$, and $\dot{Z}$, the time derivatives of the rectangular coordinates of the body position. The position of the body is obtained by integration of the Chebysev polynomials and addition of a stored constant of integration. Chebyshev polynomials of velocity are used in the EPM ephemerides. The time argument for these ephemerides uses the TDB time scale. The details of ordering of data in Type~20 kernels is given in \S\ref{Type20}.
\item Type 102. Chebyshev polynomials (TCB: position only): This type is identical to Type~2 except the time argument uses the TCB time scale rather than the TDB time scale. The details of ordering of data in Type~102 kernels is given in \S\ref{Type2}.
\item Type 103. Chebyshev polynomials (TCB: position and velocity): This type is identical to Type~3 except that the time argument  uses the TCB time scale rather than the TDB time scale. The details of ordering of data in Type~103 kernels is given in \S\ref{Type3}.
\item Type 120. Chebyshev polynomials (velocity only): This type is identical to Type~20 except the time argument  uses the TCB time scale rather than the TDB time scale. The details of ordering of data in Type~120 kernels is given in \S\ref{Type20}.
\end{itemize}

Each segment contains an arbitrary number of logical records. The total number of array addresses available by a kernel (\S\ref{DAF}) is the only restriction on length. Each record contains a set of Chebyshev coefficients valid throughout a fixed length time interval. The maximum Chebyshev polynomial degree is the same for each component and fixed for the segment; all records in the segment contain the same number of coefficients.

\subsection{The Individual Kernel Types}\label{IndividualFileTypes}
 
\subsubsection{The Segment Structure for Types 2, 3, 102, and 103}
 
\begin{figure}[h]
\centering
\includegraphics[width = \textwidth]{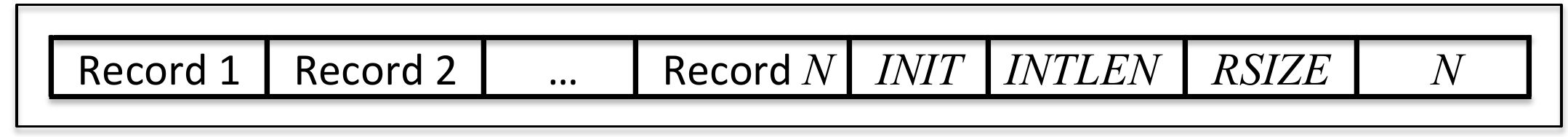}
\caption{The structure of a data segment for a Chebyshev polynomial, equal record length, SPK.}\label{SPKFig6}
\end{figure}

Figure~\ref{SPKFig6} shows the segment structure for file types 2, 3, 102, and 103. Records are ordered by increasing initial epoch. Located at the end of the segment is a {\em directory} of four numerical values. This directory contains the data required to determine the location and evaluate the record for a particular epoch:
\begin{enumerate}
\item $INIT$: The initial epoch of the first record, given in seconds from J2000.0 (JD~245\-1545.0).
\item $INTLEN$: The length of the interval covered by each record, in seconds.
\item $RSIZE$: The total number of array elements in each record.
\item $N$: The number of records contained in the segment.
\end{enumerate}
The maximum degree of the Chebyshev polynomial, $PD$, for a segment is deduced from the value of $RSIZE$.
 
The coefficients for each coordinate position or velocity in a record are stored together, in order, from the 0th to the $PD$th degree Chebyshev polynomial coefficient. If the {\em n}th degree Chebyshev polynomial is not used, its coefficient must be present and set to 0.

\subsubsection{Type 2: Chebyshev Polynomials (Position Only) and Type 102: Chebyshev Polynomials (TCB: Position Only)}\label{Type2}

The types 2 and 102 SPK data type contains Chebyshev polynomial coefficients for the position of the body as a function of time. For types 2 and 102 {\em modulo} $(RSIZE - 2, 3) = 0$, so
\begin{equation}
PD = \frac{RSIZE - 2}{3} - 1.
\end{equation}

\begin{figure}[h]
\centering
\includegraphics[width = \textwidth]{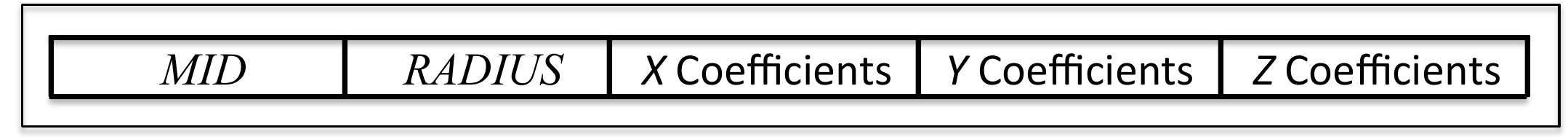}
\caption{The structure of the record for a Type~2: Chebyshev Polynomials (position only) SPK data segment.}\label{SPKFig7}
\end{figure}

Each record begins with the values for the parameters $MID$ and $RADIUS$ followed by the Chebyshev polynomial coefficients for $X$, $Y$, and $Z$ coordinate in that order. This arrangement is shown schematically in Fig.~\ref{SPKFig7}. For a Chebyshev polynomial, $Ch_n(t)$, the independent variable $t$ runs continuously from $t = -1$ to $t = 1$. $MID$ is the midpoint of the ephemeris time interval covered by the set of coefficients, $T$, in seconds from J2000.0. At $MID$ $t = 0$. And $RADIUS$, is the range, in seconds, from the midpoint to the beginning or end of the interval covered by coefficients in the record. These parameters are used to map between $T$, $MID - RADIUS \leq T \leq MID + RADIUS$, and $t$, $ -1 \leq t \leq 1$.

The difference between type 2 and type 102 segments is: The independent argument for type 2 segments is seconds from J2000.0 on the TDB time scale, while the independent argument for type 102 segments is seconds from J2000.0 on the TCB time scale. The time limits given for {\em both} types in the Segment Summary (\S\ref{SummaryRecords}), however, are given on the TDB time scale.

\subsubsection{Type 3: Chebyshev Polynomials (Position and Velocity) and Type~103: Chebyshev Polynomials (TCB: Position and Velocity) kernels}\label{Type3}

\begin{figure}[h]
\centering
\includegraphics[width = \textwidth]{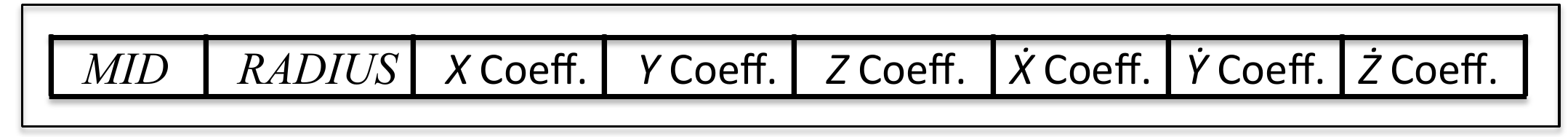}
\caption{The structure of the record for a Type~3: Chebyshev Polynomials (position and velocity) SPK data segment.} \label{SPKFig8}
\end{figure}

The types 3 and 103 SPK data type contains separate Chebyshev polynomial coefficients for the position and velocity of the body as a function of time. The structure of the segment is similar to that of the SPK data types 2 and 102. The only difference is that each logical record (see Fig.~\ref{SPKFig8}) contains six sets of coefficients instead of three. Thus, for types 3 and 103 {\em modulo} $(RSIZE - 2, 6) = 0$, so
\begin{equation}
PD = \frac{RSIZE - 2}{6} - 1.
\end{equation}

The difference between type 3 and type 103 segments is: The independent argument for type 3 segments is seconds from J2000.0 on the TDB time scale, while the independent argument for type 103 segments is seconds from J2000.0 on the TCB time scale. The time limits given for {\em both} types in the Segment Summary (\S\ref{SummaryRecords}), however, are given on the TDB time scale.

\subsubsection{Type 20: Chebyshev Polynomials (Velocity Only) and Type~120: Chebyshev Polynomials (TCB:Velocity Only)}\label{Type20}

\begin{figure}[h]
\centering
\includegraphics[width = \textwidth]{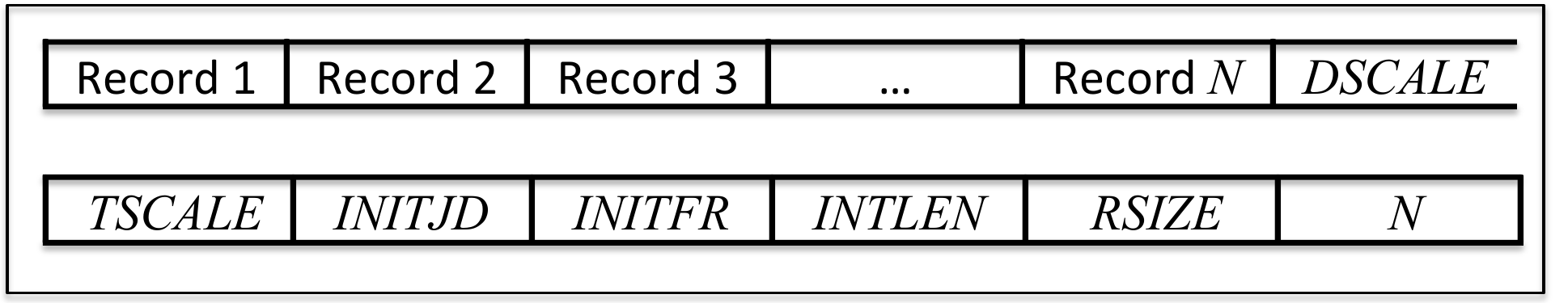}
\caption{The structure of the record for a Type~20: Chebyshev Polynomials (velocity only) SPK data segment.}\label{SPKFig85}
\end{figure}

The structure of Type 20 (Velocity Only) andType~120 (TCB:Velocity Only) data segments, displayed in Fig.~\ref{SPKFig85} are different from the other Chebyshev polynomial data segments. The parameters at the end of the data segment are:
\begin{itemize}
\item $DSCALE$: the distance scale in kilometers,
\item $TSCALE$: the time scale in seconds,
\item $INITJD$: the integer part of the Julian Date of the initial record,
\item $INITFR$: the fractional part of the Julian Date of the initial record,
\item $INTLEN$: the time period covered by each record in {\em Julian days},
\item $RSIZE$: the total number of array elements in each record
   \begin{equation}
   RSIZE = 3 \left( PD + 1 \right) 
   \end{equation}
where $PD$ is the degree of the Chebyshev polynomial, and
\item $N$: the number of records in the segment.
 \end{itemize}

The Type 20 SPK data type contains Chebyshev polynomial coefficients for the velocity of the body as a function of time plus its position at the time of the mid-point of a record. The Chebyshev polynomials are evaluated using the independent argument $t$, where $-1 \leq t \leq 1$ over each record. The value of $t$ is determined from the input time $T$ using
\begin{eqnarray}
T_{JD} & = & \frac{T}{86,400} + 2451545.0\nonumber\\
T_0    & = & T_{JD} - (INITJD + INITFR)\nonumber\\
m      & = & \frac{T_0}{INTLEN}\\
T_1    & = & T_0 - m \times INTLEN\nonumber\\
t      & = & \frac{2\;T_1}{INTLEN} - 1\nonumber
\end{eqnarray}
where $T$ is in seconds from J2000.0, $m$ is an {\em integer}, and $T_{JD}$, $T_0$, and $T_1$ are real values. 

\begin{figure}[h]
\centering
\includegraphics[width = \textwidth]{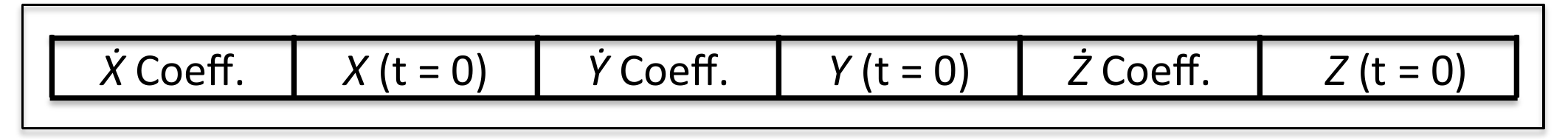}
\caption{The structure of the record for a Type 20: Chebyshev Polynomials (velocity only) SPK data segment.}\label{SPKFig9}
\end{figure}
 
Each record contains the Chebyshev polynomial coefficients for $X$, $Y$, and $Z$ velocity. The velocity coefficient for each component is followed immediately by the position component value at $t = 0$. This arrangement is shown schematically in Fig.~\ref{SPKFig9}. For a Chebyshev polynomial, $Ch_n(t)$, the independent variable $t$ runs continuously from $t = -1$ to $t = 1$. $JD(t = 0)$, the Julian Date of midpoint of a record is inferred from
\begin{equation}
JD(t = 0) = INITJD + INITFR + INTLEN\; (m - 1/2).
\end{equation}
where $(m - 1/2)$ is a {\em real} value. The units for the positions at $t = 0$ are $DSCALE$~km, and the velocities are $DSCALE / TSCALE$ km~s$^{-1}$.

The polynomial degree is fixed, so all records have the same number of parameters, so $RSIZE$ is
\begin{equation}
RSIZE = 3 \left(PD + 2 \right).
\end{equation}

The difference between type 20 and type 120 segments is: The independent argument for type 20 segments is seconds from J2000.0 on the TDB time scale, while the independent argument for type 120 segments is seconds from J2000.0 on the TCB time scale. The time limits given for {\em both} types in the Segment Summary (\S\ref{SummaryRecords}), however, are given on the TDB time scale.

\section{PCK Kernels}\label{PCKKernels}
 
PCKs are designed to supply planetary cartographic and physical constants such as planet orientation, masses, triaxial shape models, and gravitational parameters. PCK files may store either text or binary data. However, text and binary PCK files have different formats, so both text and binary data may not be stored in the same file.

Most of these data consist of a limited number of single values, short vectors, and small matrices, which are easily stored as text. NAIF supplies a text PCK \citep{Bachman2011} with the constants recommended by the IAU's Working Group on Cartographic Coordinates and Rotational Elements \citep{Archinaletal2011a, Archinaletal2011b} to  generate orientation ephemerides for major solar system bodies. The format of text PCK files is specified in \S\ref{TextPCK}.

Other supporting data are not easily stored as text. In particular, the orientations of the Earth and Moon as a function of time are complex and required to high accuracy. Thus, their orientation ephemerides are represented using Chebyshev polynomials. These polynomials may be stored in a binary PCK to both save storage and speed up evaluation. The format for binary PCKs is discussed in \S\ref{PCK}.

\subsection{Text PCKs}\label{TextPCK}

Text PCKs conform to a flexible format called ``NAIF text kernel'' format. The first line of a text PCK must consist solely of the file identification word, ``\texttt{KPL/PCK }'' starting on the first character of the line. These kernels are ASCII files designed so they may be read and modified using a text editor. Text PCKs may be used for a variety of functions, in addition to storing orientation ephemerides and supporting data described here. For more information see \citet{WrightActon2013}.

A text PCK consists of blocks (sets of contiguous lines) of comments, alternating with blocks of kernel variable assignments. A kernel variable consists of an identifying label called a ``name string'' and its value. Three kinds of data that can be placed in NAIF text kernel files:
\begin{enumerate}
\item Character strings.
\item Numerical values, stored as double precision numbers.
\item Time values, stored as seconds from J2000.0.
\end{enumerate}
Double precision number data may consist of scalars, vectors, or matrices. Values are associated with name strings using a ``name string = value'' format. The name strings, together with their associated values, are called ``kernel variables''. 
 
Comment blocks begin with the control sequence (a string alone on a line)
\begin{verse}
   \texttt{\textbackslash begintext}
\end{verse}
Data blocks begin with the control sequence
\begin{verse}
   \texttt{\textbackslash begindata}
\end{verse}
In a text kernel file, the lines preceding the first \texttt{\textbackslash{}begindata} control sequence constitute a comment block. A \texttt{\textbackslash{}begintext} control sequence is optional for this comment block.

Except for non-printing characters or lines that can be interpreted as control sequences the text of comment blocks is arbitrary.

Data blocks must appear in the form of an assignment such as
\begin{verse}
   \texttt{NAME = ( VALUE1, VALUE2, ... )}
\end{verse}
where \texttt{NAME} is a case sensitive string no longer than 32 characters. The values on the right hand side may be either numeric values or character strings. Numeric values may be either integer or floating point values. Character string values are normally limited to 80 characters in length and are single quoted. For example
\begin{verse}
   \texttt{BODY399\_RADII = (  6378.140  6378.140  6356.75  )}
\end{verse}
\begin{itemize}
\item Vector data values are separated by commas or blanks, but {\em not} by tabs.
\item The right hand side of the assignment can be continued over multiple lines.
\item Numeric values can be expressed as integers or reals.
\item Real values may be expressed in fixed point or scientific notation.
\item Scientific notation may use either ``E'' or ``D'' to delimit the exponent and is not case sensitive.
\item Vector data values are enclosed in parentheses.
\item Assignments of scalars do not require the value to be enclosed in parentheses, but that notation is frequently used as a visual cue.
\item Blank lines within or between assignments are ignored.
\item Time values may be assigned using the ``@'' character to identify it as a time value.
\end{itemize}
Non-printing characters including {\em tab} should not be present in the file: the presence of such characters may cause formatting errors when the file is viewed.

\subsubsection{Text PCK Variable Names}
 
Variable names are case-sensitive and must not exceed 32 characters in length. They may include any printable character \citep{Bachman2014} except:
\begin{itemize}
\item space, `` ''; 
\item period,``.''; 
\item parentheses,``('' or ``)'';
\item equal sign,``=''; or
\item the TAB character.
\end{itemize}
Not using the Plus sign, ``+'', as the last character is recommended.
 
Within SPICE variable names that do not have the expected case will be invisible to SPICE\-LIB routines that try to fetch their values. SPICE routines that use kernel variables accept only upper case names, so NAIF recommends upper case always be used for variable names.

Within SPICE text PCK variables follow the pattern: variables related to a body whose NAIF integer code is nnn have names of the form
\begin{verse}
   \texttt{BODY{\it nnn}\_$<$item name$>$}
\end{verse}
where \texttt{$<$item name$>$} is a short string that identifies the type of quantity the kernel variable represents. For example, the variable containing quadratic polynomial coefficients for the right ascension of the Earth's north pole is
\begin{verse}
   \texttt{BODY399\_POLE\_RA}
\end{verse}
Those not using SPICE may adopt their own convention for identifying bodies.

\subsubsection{Models for the Orientation of Solar System Bodies}

PCK kernels contain the physical and cartographic parameters required to describe extended solar system bodies. examples of data appropriate for inclusion in PCKs are radii of bodies, constants defining orientation models, masses or values of GM.

For the Sun, planets, and satellites, these parameters are denoted by:
\begin{itemize}
\item \texttt{BODY{\em nnn}\_GM} -- The mass times the gravitational constant in km$^3$~s$^{-2}$.
\item \texttt{BODY{\em nnn}\_POLE\_RA} -- The right ascension of the pole of orientation\footnote{For the Sun, planets and satellites, the pole of orientation is the north pole, defined as the pole on the north side of the invariable plane of the solar system. For dwarf planets, minor planets, and comets the pole of orientation is the pole around which the body rotates in a counterclockwise direction \citep{Archinaletal2011a}.\label{polefn}}.
\item \texttt{BODY{\em nnn}\_POLE\_DEC} -- The declination of the pole of orientation.
\item \texttt{BODY{\em nnn}\_PM} -- The prime meridian location\footnote{The position of the prime meridian is measured in the counterclockwise direction, from the vector resulting from the cross product of the unit vector pointing towards the north pole of the ICRS and pointing towards the body's pole of orientation at time $t$ \citep{Archinaletal2011a}.}.
\item \texttt{BODY{\em nnn}\_NUT\_PREC\_RA} -- The amplitudes of the components of nutation of the pole in right ascension.
\item \texttt{BODY{\em nnn}\_NUT\_PREC\_DEC} -- The amplitudes of the components of nutation of the pole in declination.
\item \texttt{BODY{\em nnn}\_NUT\_PREC\_PM} -- The amplitudes arising from the components of nutation to the rotation of the body.
\item \texttt{BODY{\em bbb}\_NUT\_PREC\_ANGLES} -- The value at epoch and time rate of change of the body nutation angles (See below).
\end{itemize}
The value of {\em nnn} is the NAIF identification number for the body whose orientation is being described. The value of {\em bbb} in \texttt{BODY{\em bbb}\_NUT\_PREC\_ANGLES} is that of the body-satellite system barycenter for which the nutation angles are valid.

\paragraph{Body Orientation:} The orientation of solar system bodies is represented by three angles: $\alpha$ and $\delta$ are the right ascension and declination of a body's pole of orientation\footnote{See footnote~\ref{polefn}.}, and $W$ is the prime meridian location angle, measured in the counterclockwise direction about the body's pole of orientation, from the vector resulting from the cross product of the unit vector pointing towards the north pole of the ICRS and the unit vector pointing towards the body's pole of orientation at time $t$. The expressions used in text PCK files for the direction of the pole of orientation and prime meridian location consists of the sum of a quadratic polynomial part and a periodic part. The general form is
\begin{equation}
X = X_0 + \frac{X_1\;t}{T} + \frac{X_2\;t^2}{T^2} + \sum_i x_i \left\{ \begin{array}{ll}
    \sin \theta_i & \mbox{ for $\alpha$ and $W$}\\
    \cos \theta_i & \mbox{ for $\delta$}
\end{array} \right.
\end{equation}
where $t$ is the ephemeris time in seconds from the reference epoch, $X_0$, $X_1$, and $X_2$, are the coefficients for the polynomial portion of the body's motion, $x_i$ is the array of coefficients for the periodic portion, and $\theta_i$ is the array of periodic angles.
\begin{itemize}
\item For $\alpha$ and $\delta$ the polynomial portion represents the secular precession of the pole of the body, and the periodic part represents the nutation of the pole. The value of $T = 3,155,760,000$, the number of seconds in a Julian century.
\item For $W$, the polynomial represents the rotation of the body with respect to the ICRS and the periodic part represents librations in the rotation. The value of $T = 86,400$, the number of seconds in a day.
\end{itemize}

\paragraph{The Nutation/Libration Angles:} The nutation/libration angles, $\theta_i$, which form the periodic portion of solar system bodies, arise from the spin-orbit coupling between the body of interest and other solar system bodies. Usually, only the torque of the central body and satellites of a body-satellite system are large enough to be significant. Thus, their values are given using the body-satellite barycenter as the reference. The angles are represented by $\theta_{i0}$, the value of $\theta_i$ at the epoch J2000.0, and $\theta_{i1}$, its time rate of change. The value of $\theta_i$, in degrees, at time $t$ is given by
\begin{equation}
\theta_i = \theta_{i0} + \frac{\theta_{i1}\;t}{T}
\end{equation}
where the value of $T = 3,155,760,000$, the number of seconds in a Julian century.

All of the nutation angles for a body-satellite system may not apply to all of the orientation parameters. If the $i$th periodic component does not affect a parameter, its amplitude $x_i = 0$.

\subsubsection{The Radii of Solar System Bodies}

The radii of solar system bodies are given as a triplet to allow for triaxial shapes. The naming convention is \texttt{BODY{\em nnn}\_RADII}. The radii are those of the body-fixed $X$, $Y$, and $Z$-axes in units of kilometers.

\subsubsection{Epoch and Frame Specification}

By default, the orientation model parameters used in PCKs are assumed to define the transformation of vectors expressed in a base reference frame (normally the ICRS) to a body-fixed reference frame. The orientation of the latter frame is evaluated at $t$ seconds from the epoch J2000.0. This transformation is given in the form of a rotation, $R(t)$. On those rare occasions where needed, the default values for the epoch and frame of the constants may be overridden. For example, it is possible to use constants referenced to the B1950 frame or the J1950 epoch.

The default reference frame and reference epoch for a body are overridden by setting the values of the kernel variables
\begin{verse}
   \texttt{BODY{\em nnn}\_CONSTANTS\_REF\_FRAME}
\end{verse}
and
\begin{verse}
   \texttt{BODY{\em nnn}\_CONSTANTS\_JED\_EPOCH}
\end{verse}
where {\em nnn} is the NAIF identification number for the planet-satellite barycenter or, for other objects, the identification number of the body itself.

The values of the frame specifier variable
\begin{verse}
   \texttt{BODY{\em nnn}\_CONSTANTS\_REF\_FRAME}
\end{verse}
are the codes for the reference frames recognized by SPICE (\S\ref{RecognizedReferenceFrames}).

For example, to use constants referenced to the FK4 frame for the asteroid Gaspra (Body Identification Number~=~9511010), a text PCK containing the constants should include the assignment
\begin{verse}
   \texttt{BODY9511010\_CONSTANTS\_REF\_FRAME   =   (  3  )}
\end{verse}

The values of the epoch specifier variable
\begin{verse}
   \texttt{BODY{\em nnn}\_CONSTANTS\_JED\_EPOCH}
\end{verse}
are Julian ephemeris dates. To use constants for Gaspra referenced to the J1950.0 epoch, a text PCK containing the constants should include the assignment
\begin{verse}
   \texttt{BODY9511010\_CONSTANTS\_JED\_EPOCH   =   ( 2433282.5D0 )}
\end{verse}

The same frame and epoch must be used for each planet-satellite system, but the frame and epoch of the constants for each system or other body may be set independently. For example, to reference the Earth-Moon system to the B1950 frame and J1950 epoch the assignments are
\begin{verse}
   \texttt{BODY3\_CONSTANTS\_REF\_FRAME   =   (  2           )\\
   BODY3\_CONSTANTS\_JED\_EPOCH   =   (  2433282.5D0 )}
\end{verse}
where the $nnn = 3$ designates the Earth-Moon barycenter. The assignment
\begin{verse}
   \texttt{BODY399\_CONSTANTS\_REF\_FRAME   =   (  2           )\\
   BODY399\_CONSTANTS\_JED\_EPOCH   =   (  2433282.5D0 )}
\end{verse}
would be ignored by the SPICE text PCK reader routines, since a frame or epoch cannot be assigned to a planet or satellite.

\subsubsection{Example}

The following example is an edited snippet of the 2011~October SPICE solar system body orientation file pck00010.tpc \citep{Bachman2011}.

\begin{verse}
\texttt{KPL/PCK \\
\medskip
P\_constants (PCK) SPICE kernel file\\
\medskip
Purpose\\
--------------------------------------------------------\\
\medskip
This file makes available for use in SPICE-based applications\\
software orientation and size/shape data for natural bodies. The\\
principal source of the data is a published report by the IAU\\
Working Group on Cartographic Coordinates and Rotational Elements.\\
\medskip
Mercury\\
\medskip
\textbackslash begindata\\
\medskip
\hspace{32pt} BODY199\_POLE\_RA \hspace{1pt} = ( 281.0097 -0.0328 \hspace{10pt} 0. )\\
\hspace{32pt} BODY199\_POLE\_DEC  = ( \hspace{1pt} 61.4143 -0.0049 \hspace{10pt}  0. )\\
\hspace{32pt} BODY199\_PM \hspace{24pt} = (  329.5469 \hspace{1pt} 6.1385025  0. )\\
\medskip
\hspace{32pt} BODY199\_NUT\_PREC\_RA \hspace{1pt} = (  0. 0. 0. 0. 0. )\\
\hspace{32pt} BODY199\_NUT\_PREC\_DEC = (  0. 0. 0. 0. 0. )\\
\hspace{32pt} BODY199\_NUT\_PREC\_PM \hspace{1pt} = (  0.00993822\\
\hspace{148pt} -0.00104581\\
\hspace{148pt} -0.00010280\\
\hspace{148pt} -0.00002364\\
\hspace{148pt} -0.00000532 )\\
\medskip
\textbackslash begintext\\
\medskip
The linear coefficients have been scaled up from degrees/day\\
to degrees/century, because the SPICE PCK reader expects\\
these units.  The original constants were:\\
\medskip
\textbackslash begindata\\
\medskip
\hspace{32pt} BODY1\_NUT\_PREC\_ANGLES = (174.791086  0.14947253587500003E+06\\
\hspace{153pt} 349.582171  0.29894507175000006E+06\\
\hspace{153pt} 164.373257  0.44841760762500006E+06\\
\hspace{153pt} 339.164343  0.59789014350000012E+06\\
\hspace{153pt} 153.955429  0.74736267937499995E+06)\\
\medskip
\textbackslash begintext\\
\medskip
===================================================================\\
End of file pck00010.tpc\\
===================================================================}
\end{verse}

\subsection{The Binary PCK Format}\label{PCK}

Binary PCKs provide the orientation of a body-fixed reference frame, with respect to an inertial reference frame, sometimes called the ``base frame''. This property of binary PCKs may be used to store the orientation of bodies whose models are too complicated to be economically stored as a set of coefficients. The only natural solar system bodies whose orientations are currently well enough known to require the use of a binary PCK file are the Earth and the Moon. The binary PCK format uses the DAF (\S\ref{DAF}) architecture, and its structure is similar to that of the SPK. The DAF file identification word occupying the first eight bytes of a binary PCK is ``\texttt{DAF/PCK }''.

Like SPKs, the binary PCK summary has two double precision components ($ND = 2$), and five integer components ($NI = 5$). The double precision components of the summary are:
\begin{enumerate}
\item the initial epoch and
\item the final epoch
\end{enumerate}
of the interval for which data are contained in the segment, in seconds using the TDB time scale from J2000.0 (JD~$245\:1545.0$). The initial and final epochs are always given using the TDB time scale even when the time argument used in the segment is TCB. The integer components of the summary are:
\begin{enumerate}
\item the NAIF body-fixed frame class identification number (The reference frame for which the ephemeris describes its orientation.);
\item the NAIF base frame identification number (The inertial reference frame with respect to which the body-fixed frame ephemeris is referenced.);
\item the integer code for the representation (type PCK data: \S\ref{BinaryPCK});
\item the initial address of the array; and
\item the final address of the array.
\end{enumerate}
The first two integer components of the summary are discussed in \S\ref{BinaryPCKFrames}, the third integer component is discussed in \S\ref{BinaryPCK} and the last last two integer components are discussed in \S\ref{ArrayAddr}.

The Coordinate Time ephemeris identification numbers ($1\,000\,000\,001$,
$1\,000\,000\,002$, and $1\,000\,\-000\,003$) are {\em not} recognized by PCKs.

\subsubsection{Binary PCK Reference Frames}\label{BinaryPCKFrames}

The rotation of a body is the equivalent of the change in orientation of one reference frame, for example the axes of the mean lunar principal moments of inertia, with respect to a second reference frame, for example the ICRS, as a function of time. The first two integer components of the binary PCK summary are the NAIF frame class identification number of the body-fixed frame and the NAIF frame  identification number\footnote{Not the frame {\em class} identification number.} of the inertial base frame. The base frame {\em must} be an inertial reference frame.

The identification number for the ICRS is 1. Other recognized inertial reference frame names and their identity numbers that might be of interest are in Table~\ref{InertialFramesTable}.

SPICE contains a number of built-in non-inertial reference frame names and identification numbers \citep{Semenov2010b}. The only non-inertial reference frames that would be of interest for binary PCKs are the Earth and the Moon. These reference frames are summarized in Table~\ref{NonInertialFramesTable}.

NAIF makes available the ITRF93 Earth orientation as a binary PCK on its web site\footnote{ftp://naif.jpl.nasa.gov/pub/naif/generic\_kernels/pck}. The frame identification number is 13000. \citet{Bachman2007} gives details.

The situation for the Moon is more complicated. The Moon has two predominant reference frames: the axes of the principal moments of inertia (PA) reference frame and the mean Earth-mean pole of rotation (ME) reference frame. The PA reference frame is used for dynamical analyses of the Moon because its use simplifies the equations of motion. The ME reference frame is the historical reference frame used for locating features on the lunar surface. The difference between these two reference frames is a constant rotation. The ME reference frame is not stored as a binary PCK. It is stored as a rotation with respect to the PA reference frame in a text reference frame kernel\footnote{The format of a text reference frame kernel is similar to that of a text PCK. The main difference is that the identification word occupying the first eight bytes of a text reference kernel are ``\texttt{KPL/FK  }''.}. Furthermore, the precise orientation of the axes of the principal moments of inertia cannot cleanly be separated from other lunar model parameters with the available data \citep{Williamsetal2013}. Hence, both the PA and ME reference frames are ephemeris dependent. To deal with this model dependence each ephemeris solution is given a unique reference frame identification number. These ephemerides are then linked to the generic reference frame numbers by SPICE. For example the DE421 PA reference frame name and number are MOON\_PA\_DE421 and 31006, and the ME reference frame name and number are MOON\_ME\_DE421 and 31007. The data providing the PA and MA reference frame identification names and numbers are part of the text reference frame kernel data. For details see \citet{Bachman2008}.

\subsubsection{Binary PCK Data Types}\label{BinaryPCK}

The third integer component of the summary is the code for the representation, or {\em data type}. The data type plays no role in selecting the segment to satisfy a data request. The type used to represent the data becomes important only when the data in a segment are to be evaluated. This step is isolated, so new data types can be added to the binary PCK format without affecting application programs that use the higher level readers.

SPICE currently implements two binary PCK data types that use Chebyshev polynomials. They are:
\begin{itemize}
\item Type 2. Chebyshev polynomials (Angles): These are sets of coefficients for the angular components of the body orientation. This type is identical in structure to the Type 2 SPK; details of ordering of data are discussed in \S\ref{PCKType2}.
\item Type 20. Chebyshev polynomials (Angle Rates): These are sets of coefficients for the time rate of change of the angular components of the body orientation. This type is identical in structure to the Type 20 SPK; details of ordering of data are discussed in \S\ref{PCKType20}.
\end{itemize}
Kernels that do not use the correct format for these data types will not be correctly interpreted by SPICE. Two other binary PCK data types have been defined, but are not implemented in SPICE. They are:
\begin{itemize}
\item Type 102. Chebyshev polynomials (TCB: Angles): These are sets of coefficients for the angular components of the body orientation. This type is identical in structure to the Type~102 SPK; details of ordering of data are discussed in \S\ref{PCKType2}.
\item Type 120. Chebyshev polynomials (TCB: Angle Rates): These are sets of coefficients for the time rate of change of the angular components of the body orientation where the time argument uses the TCB time scale. This type is identical in structure to the Type~120 SPK; details of ordering of data are discussed in \S\ref{PCKType20}.
\end{itemize}
PCK data type identifiers 901 through 910 have been set aside for user defined data types. NAIF may also reserve other data types in the future. 
 
\subsubsection{Type 2: Chebyshev Polynomials (Angles) and Type~102: Chebyshev Polynomials (TCB: Angles)}\label{PCKType2}

PCK type 2 data type segments contain Chebyshev polynomial coefficients for the  orientation of a body as a function of time. The structure is identical to the Type~2 SPK data type.

Each record begins with the values for the parameters $MID$ and $RADIUS$ followed by the Chebyshev polynomial coefficients for angles $\phi$, $\theta$, and $\psi$, where, in SPICE, $\phi$ is the angle in the $X-Y$ plane of the inertial reference frame with its apex at the origin of the body's reference frame from the positive $X$-axis of the inertial reference frame to the ascending node of the body's $X-Y$ plane, $\theta$ is the inclination of the body's $X-Y$ plane to the inertial reference frame's $X-Y$ plane, and $\psi$ is the angle in the body's $X-Y$ plane from the ascending node to the body's positive $X$-axis. The arrangement of data is shown schematically in Fig.~\ref{SPKFig7}, replacing $X$, $Y$, and $Z$ with $\phi$, $\theta$, and $\psi$, respectively. Like the Type~2 SPK data type, the polynomial degree is fixed, so all records have the same number of parameters, and $RSIZE$ is
\begin{equation}
RSIZE = 3 \left(PD + 1 \right) + 2
\end{equation}
where $PD$ is the polynomial degree.

PCK type 102 segments are identical to PCK type~2 data segments except that the independent variable uses the TCB time scale rather than the TDB time scale.

\subsubsection{Type 20: Chebyshev Polynomials (Angle Rates) and Type~120: Chebyshev Polynomials (TCB: Angle Rates) kernels}\label{PCKType20}

PCK type 20 PCK data segments contain separate Chebyshev polynomial coefficients for the rate of change in orientation of the body and its orientation at $t = 0$ as a function of time. The structure of the segment is identical to that of the SPK data Type~20: Chebyshev Polynomials (velocity only). The only difference is that each logical record (see Fig.~\ref{SPKFig8}) contains coefficients for the time derivatives of the angles $\phi$, $\theta$, and $\psi$ instead of velocities.

Like the Type 20 SPK data type, the polynomial degree, $PD$, is fixed, so the same number of coefficients is always used for each component, and all records are the same size:
\begin{equation}
RSIZE = 3 \left(PD + 1 \right) + 2.
\end{equation}

PCK type 120 data segments are identical to PCK type~20 data segments except that the independent variable uses the TCB time scale rather than the TDB time scale.

\section{Units}\label{Units}

In general, the SPICE Toolkit assumes that for SPK and PCK data the unit of  length is the kilometer, the unit of angle is the radian, and the unit of time is the second. There are some exceptions the user needs to be aware of:
\begin{itemize}
\item PCK types 3 and 103: Angular units are degrees and angular rates are    degrees~second$^{-1}$.
\item SPK type 20 and 120 (\S\ref{Type20}) and PCK types 20 and 120 (\S\ref{PCKType20}): Distance and time units are set by the file's creator.
\end{itemize}
 
\section{Validation}
 
Validating correct formatting is essential for complex SPK and PCK kernels. Consider writing an application that compares states from source data with states extracted from the kernel. Include states interpolated from source data not used in generating states placed in the kernel. Verify a uniform fit over the full time interval covered by the kernel.

Testing with SPK or PCK kernels with a known provenance are read successfully is one method to assist in validation. Examples of text PCK kernels, binary PCK type 2 and SPK type 2 kernels are available from the NAIF web site\footnote{http://naif.jpl.nasa.gov/pub/naif/generic\_kernels/} and the INPOP web site\footnote{http://www.imcce.fr/inpop/download13c.php} at the IMCCE. Examples of text PCK kernels, binary PCK type 20 and SPK type 20 kernels are available from the IAA web site\footnote{ftp://quasar.ipa.nw.ru/incoming/EPM/Data/}.

The utility applications SPY, for making structure and semantic checks of kernels, and SPKDIFF, for comparing similar kernels, are available at the NAIF web site\footnote{http://naif.jpl.nasa.gov/naif/utilities.html\label{NAIFWebRef}}. These applications can assist in validating kernels.

One common problem with kernels is time gaps between segments. Time gaps, even sub-second gaps, may cause problems for users.

Occasionally, merging portions of two or more SPK kernels into one is required. This may be done by writing an appropriate application. The SPICE utility application SPKMERGE is available from NAIF\footref{NAIFWebRef}. Carefully examine the results to verify the result is what was expected. The NAIF utility applications BRIEF and SPACIT are available to help with the verification. Verify that the comments in the merged kernel are appropriate.

 

\bibliographystyle{spbasic}      

\appendix
\clearpage

\section{NAIF Body Identification Numbers}\label{ObjectIDNosApp}

Each object in the solar system is assigned a unique NAIF identification number. This appendix explains how these numbers are assigned, and the tables give the available identification numbers recognized by SPICE at the time this report was written. Current tables of identification numbers may be found at the NAIF web site \citep{Wright2010}.

\subsection{The Sun and Planetary Barycenters}\label{Barycenters}

The smallest positive identification numbers are reserved for planetary barycenters and the Sun. {\em These values are the ones most likely to be used for planetary ephemerides.} The NAIF identification numbers for these objects are given in Table~\ref{BarycenterIDs}. For those planets without satellites, Mercury and Venus, the barycenter location is identical to the location of the body center of mass. However, a planet barycenter identification number may not be interchanged with a planet identification number described in \S\ref{PlanetsandSatellites}. A barycenter has only the attributes of mass and position, while a planet has additional attributes such as size, shape and rotation pole and rate of rotation. The position ephemeris of a planet-satellite system is designated by its barycenter identification number, but the physical ephemeris of the planet itself is designated by its planet identification number.

\begin{table}[h]
\begin{center}
\caption{NAIF identification numbers for the Sun and planetary barycenters.\label{BarycenterIDs}}
\begin{tabular}{c l}
\hline\hline
NAIF ID & \multicolumn{1}{c}{Name} \\
\hline
 0 & Solar System Barycenter\\
 1 & Mercury Barycenter\\
 2 & Venus Barycenter\\
 3 & Earth-Moon Barycenter\\
 4 & Mars Barycenter\\
 5 & Jupiter Barycenter\\
 6 & Saturn Barycenter\\
 7 & Uranus Barycenter\\
 8 & Neptune Barycenter\\
 9 & Pluto$^a$ Barycenter\\
10 & Sun\\
\hline
\multicolumn{2}{l}{$^a$Pluto is included here due to its pre-}\\ 
\multicolumn{2}{l}{vious classification as a planet.}\\
\end{tabular}
\end{center}
\end{table}

\clearpage

\subsection{Planets and Satellites}\label{PlanetsandSatellites}

The format of planet identification numbers is:
\begin{eqnarray}
P99\nonumber
\end{eqnarray}
where $P$ is the number of the planet in increasing distance from the Sun, {\em e.g.}~Jupiter is body number 599. Pluto is included in this scheme as a continuation of its previous status as a planet.

Similarly, the format for satellites identification numbers is either
\begin{eqnarray}
PNN & {\rm or} & PXNNN\nonumber
\end{eqnarray}
where $N$ is a digit and $X$ is 0 or 5. The digits $NN$ or $NNN$ are unique for a given value of $P$. These digits are the same as the IAU Roman numerals for a satellite. For example, Ananke, the 12th satellite of Jupiter (JXII), is body number 512. Codes with $X = 5$ are provisional.
 
The current planet and satellite identification numbers recognized by SPICE \citep{Wright2010} are given in Table~\ref{PlanetSatIDs}.

\begin{longtable}{c l l c l l}
\caption{NAIF identification numbers for planets and satellites.\label{PlanetSatIDs}}\\
\hline\hline
NAIF ID & Name &IAU Desig. & NAIF ID & Name & IAU Desig.\\
\hline
\endfirsthead
\caption{(Continued)}\\
\hline\hline
NAIF ID & Name &IAU Desig. & NAIF ID & Name & IAU Desig.\\
\hline
\endhead
199 & Mercury    &           & 516 & Metis      & JXVI    \\
    &            &           & 517 & Callirrhoe & JXVII   \\
299 & Venus      &           & 518 & Themisto   & JXVIII  \\
    &            &           & 519 & Magaclite  & JXIX    \\
399 & Earth      & EI        & 520 & Taygete    & JXX     \\
301 & Moon       &           & 521 & Chaldene   & JXXI    \\
    &            &           & 522 & Harpalyke  & JXXII   \\
499 & Mars       &           & 523 & Kalyke     & JXXIII  \\
401 & Phobos     & MI        & 524 & Iocaste    & JXXIV   \\
402 & Deimos     & MII       & 525 & Erinome    & JXXV    \\
    &            &           & 526 & Isonoe     & JXXVI   \\
599 & Jupiter    &           & 527 & Praxidike  & JXXVII  \\
501 & Io         & JI        & 528 & Autonoe    & JXXVIII \\
502 & Europa     & JII       & 529 & Thyone     & JXXIX   \\
503 & Ganymede   & JIII      & 530 & Hermippe   & JXXX    \\
504 & Callisto   & JIV       & 531 & Aitne      & JXXXI   \\
505 & Amalthea   & JV        & 532 & Eurydome   & JXXXII  \\
506 & Himalia    & JVI       & 533 & Euanthe    & JXXXIII \\
507 & Elara      & JVII      & 534 & Euporie    & JXXXIV  \\
508 & Pasiphae   & JVIII     & 535 & Orthosie   & JXXXV   \\
509 & Sinope     & JIX       & 536 & Sponde     & JXXXVI  \\
510 & Lysithea   & JX        & 537 & Kale       & JXXXVII \\
511 & Carme      & JXI       & 538 & Pasithee   & JXXXVIII\\
512 & Ananke     & JXII      & 539 & Hegemone   & JXXXIX  \\
513 & Leda       & JXIII     & 540 & Mneme      & JXL     \\
514 & Thebe      & JXIV      & 541 & Aoede      & JXLI    \\
515 & Adrastea   & JXV       & 542 & Thelxinoe  & JXLII   \\
 
543 & Arche      & JXLIII    & 635 & Daphnis    & SXXXV   \\
544 & Kallichore & JXLIV     & 636 & Aegir      & SXXXVI  \\
545 & Helike     & JXLV      & 637 & Bebhionn   & SXXXVII \\
546 & Carpo      & JXLVI     & 638 & Bergelmir  & SXXXVIII\\
547 & Eukelade   & JXLVII    & 639 & Bestla     & SXXXIX  \\
548 & Cyllene    & JXLVIII   & 640 & Farbauti   & SXL     \\
549 & Kore       & JXLIX     & 641 & Fenrir     & SXLI    \\
550 & Herse      & JL        & 642 & Fornjot    & SXLII   \\
    &            &           & 643 & Hati       & SXLIII  \\
699 & Saturn     &           & 644 & Hyrokkin   & SXLIV   \\
601 & Mimas      & SI        & 645 & Kari       & SXLV    \\
602 & Enceladus  & SII       & 646 & Loge       & SXLVI   \\
603 & Tethys     & SIII      & 647 & Skoll      & SXLVII  \\
604 & Dione      & SIV       & 648 & Surtur     & SXLVIII \\
605 & Rhea       & SV        & 649 & Anthe      & SXLIX   \\
606 & Titan      & SVI       & 650 & Jarnsaxa   & SL      \\
607 & Hyperion   & SVII      & 651 & Greip      & SLI     \\
608 & Iapetus    & SVIII     & 652 & Tarqeq     & SLII    \\
609 & Phoebe     & SIX       & 653 & Aegaeon    & SLIII   \\
610 & Janus      & SX        &     &            &         \\
611 & Epimetheus & SXI       & 799 & Uranus     &         \\
612 & Helene     & SXII      & 701 & Ariel      & UI      \\
613 & Telesto    & SXIII     & 702 & Umbriel    & UII     \\
614 & Calypso    & SXIV      & 703 & Titania    & UIII    \\
615 & Atlas      & SXV       & 704 & Oberon     & UIV     \\
616 & Prometheus & SXVI      & 705 & Miranda    & UV      \\
617 & Pandora    & SXVII     & 706 & Cordelia   & UVI     \\
618 & Pan        & SXVIII    & 707 & Ophelia    & UVII    \\
619 & Ymir       & SXIX      & 708 & Bianca     & UVIII   \\
620 & Paaliaq    & SXX       & 709 & Cressida   & UIX     \\ 
621 & Tarvos     & SXXI      & 710 & Desdemona  & UX      \\
622 & Ijiraq     & SXXII     & 711 & Juliet     & UXI     \\
623 & Suttungr   & SXXIII    & 712 & Portia     & UXII    \\
624 & Kiviuq     & SXXIV     & 713 & Rosalind   & UXIII   \\
625 & Mundilfari & SXXV      & 714 & Belinda    & UXIV    \\
626 & Albiorix   & SXXVI     & 715 & Puck       & UXV     \\
627 & Skathi     & SXXVII    & 716 & Caliban    & UXVI    \\
628 & Erriapus   & SXXVIII   & 717 & Sycorax    & UXVII   \\
629 & Siarnaq    & SXXIX     & 718 & Prospero   & UXVIII  \\
630 & Thrymr     & SXXX      & 719 & Setebos    & UXIX    \\
631 & Narvi      & SXXXI     & 720 & Stephano   & UXX     \\
632 & Methone    & SXXXII    & 721 & Trinculo   & UXXI    \\
633 & Pallene    & SXXXIII   & 722 & Francisco  & UXXII   \\
634 & Polydeuces & SXXXIV    & 723 & Margaret   & UXXIII  \\
  
724 & Ferdinand  & UXXIV     & 807 & Larissa    & NVII    \\
725 & Perdita    & UXXV      & 808 & Proteus    & NVIII   \\
726 & Mab        & UXXVI     & 809 & Halimede   & NIX     \\
727 & Cupid      & UXXVII    & 810 & Psamathe   & NX      \\
    &            &           & 811 & Sao        & NXI     \\
899 & Neptune    &           & 812 & Laomedeia  & NXII    \\
801 & Triton     & NI        & 813 & Neso       & NXIII   \\
802 & Nereid     & NII       &     &            &         \\
803 & Naiad      & NIII      & 999 & Pluto$^a$  &         \\
804 & Thalassa   & NIV       & 901 & Charon     & PI      \\
805 & Despina    & NV        & 902 & Nix        & PII     \\
806 & Galatea    & NVI       & 903 & Hydra      & PIII    \\          
\hline
\multicolumn{6}{l}{$^a$Pluto is included here because of its previous classification as a planet.}
\end{longtable}

\clearpage

\subsection{Comets}

Identification numbers for periodic comets begin at 1,000,001 and continue in sequence up to 2,000,\-000. The current list of periodic comets \citep{Wright2010} and their identification numbers are given in Table~\ref{CometIDs}. The identification number for a new comet is formed by adding one to the last comet identification number in the current SPICE list. The first part of the list through identification number 1,000,112 is in alphabetical order. Comet Shoemaker Levy 9 is included in this list, identification number 1,000,130, though it is no longer a comet, periodic or otherwise.

\begin{longtable}{c l c l}
\caption{NAIF identification numbers for comets.}\label{CometIDs}\\
\hline\hline
NAIF ID & Name & NAIF ID & Name\\
\hline
\endfirsthead
\caption{(Continued).}\\
\hline\hline
NAIF ID & Name & NAIF ID & Name\\
\hline
\endhead
1000001 & Arend                    & 1000037 & Haneda-Campos          \\
1000002 & Arend-Rigaux             & 1000038 & Harrington             \\
1000003 & Ashbrook-Jackson         & 1000039 & Harrington-Abell       \\
1000004 & Boethin                  & 1000040 & Hartley 1              \\
1000005 & Borrelly                 & 1000041 & Hartley 2              \\
1000006 & Bowell-Skiff             & 1000042 & Hartley-IRAS           \\
1000007 & Bradfield                & 1000043 & Herschel-Rigollet      \\
1000008 & Brooks 2                 & 1000044 & Holmes                 \\
1000009 & Brorsen-Metcalf          & 1000045 & Honda-Mrkos-Pajdusakova\\
1000010 & Bus                      & 1000046 & Howell                 \\
1000011 & Chernykh                 & 1000047 & IRAS                   \\
1000012 & Churyumov-Gerasimenko    & 1000048 & Jackson-Neujmin        \\
1000013 & Ciffreo                  & 1000049 & Johnson                \\
1000014 & Clark                    & 1000050 & Kearns-Kwee            \\
1000015 & Comas Sola               & 1000051 & Klemola                \\
1000016 & Crommelin                & 1000052 & Kohoutek               \\
1000017 & d'Arrest                 & 1000053 & Kojima                 \\
1000018 & Daniel                   & 1000054 & Kopff                  \\
1000019 & de Vico-Swift-NEAT       & 1000055 & Kowal 1                \\
1000020 & Denning-Fujikawa         & 1000056 & Kowal 2                \\
1000021 & du Toit 1                & 1000057 & Kowal-Mrkos            \\
1000022 & du Toit-Hartley          & 1000058 & Kowal-Vavrova          \\
1000023 & du Toit-Neujmin-Delporte & 1000059 & Longmore               \\
1000024 & Dubiago                  & 1000060 & Lovas 1                \\
1000025 & Encke                    & 1000061 & Machholz               \\
1000026 & Faye                     & 1000062 & Maury                  \\
1000027 & Finlay                   & 1000063 & Neujmin 1              \\
1000028 & Forbes                   & 1000064 & Neujmin 2              \\
1000029 & Gehrels 1                & 1000065 & Neujmin 3              \\
1000030 & Gehrels 2                & 1000066 & Olbers                 \\
1000031 & Gehrels 3                & 1000067 & Peters-Hartley         \\
1000032 & Giacobini-Zinner         & 1000068 & Pons-Brooks            \\
1000033 & Giclas                   & 1000069 & Pons-Winnecke          \\
1000034 & Grigg-Skjellerup         & 1000070 & Reinmuth 1             \\
1000035 & Gunn                     & 1000071 & Reinmuth 2             \\
1000036 & Halley                   & 1000072 & Russell 1              \\
 
1000073 & Russell 2                & 1000103 & van Houten-Lemmon      \\
1000074 & Russell 3                & 1000104 & West-Kohoutek-Ikemura  \\
1000075 & Russell 4                & 1000105 & Whipple                \\
1000076 & Sanguin                  & 1000106 & Wild 1                 \\
1000077 & Schaumasse               & 1000107 & Wild 2                 \\
1000078 & Schuster                 & 1000108 & Wild 3                 \\
1000079 & Schwassmann-Wachmann 1   & 1000109 & Wirtanen               \\
1000080 & Schwassmann-Wachmann 2   & 1000110 & Wolf                   \\
1000081 & Schwassmann-Wachmann 3   & 1000111 & Wolf-Harrington        \\
1000082 & Shajn-Schaldach          & 1000112 & Lovas 2                \\
1000083 & Shoemaker 1              & 1000113 & Urata-Niijima          \\
1000084 & Shoemaker 2              & 1000114 & Wiseman-Skiff          \\
1000085 & Shoemaker 3              & 1000115 & Helin                  \\
1000086 & Singer Brewster          & 1000116 & Mueller                \\
1000087 & Slaughter-Burnham        & 1000117 & Shoemaker-Holt 1       \\
1000088 & Smirnova-Chernykh        & 1000118 & Helin-Roman-Crockett   \\
1000089 & Stephan-Oterma           & 1000119 & Hartley 3              \\
1000090 & Swift-Gehrels            & 1000120 & Parker-Hartley         \\
1000091 & Takamizawa               & 1000121 & Helin-Roman-Alu 1      \\
1000092 & Taylor                   & 1000122 & Wild 4                 \\
1000093 & Tempel 1                 & 1000123 & Mueller 2              \\
1000094 & Tempel 2                 & 1000124 & Mueller 3              \\
1000095 & Tempel-Tuttle            & 1000125 & Shoemaker-Levy 1       \\
1000096 & Tritton                  & 1000126 & Shoemaker-Levy 2       \\
1000097 & Tsuchinshan 1            & 1000127 & Holt-Olmstead          \\
1000098 & Tsuchinshan 2            & 1000128 & Metcalf-Brewington     \\
1000099 & Tuttle                   & 1000129 & Levy                   \\
1000100 & Tuttle-Giacobini-Kresak  & 1000130 & Shoemaker-Levy 9       \\
1000101 & Vaisala 1                & 1000131 & Hyakutake              \\
1000102 & Van Biesbroeck           & 1000132 & Hale-Bopp              \\
\hline
\end{longtable}

\subsection{Asteroids and Dwarf Planets}

The format for identification numbers of numbered asteroids and dwarf planets, aside from Pluto, is:
\begin{eqnarray}
      {\rm identification\; number} & = & 2,000,000 + {\rm IAU\; asteroid\; number}\nonumber
\end{eqnarray}
For example, asteroid (2956)~Yeomans has identification number 2,002,956.
  
Due to its previous classification as a planet, Pluto and its satellites are included with the planetary barycenters (\S\ref{Barycenters}) and with planetary centers and their satellites (\S\ref{PlanetsandSatellites}).

There are three other exceptions to the asteroid identification assignment rule
\begin{itemize}
\item (951)~Gaspra: identification number $9\,511\,010$
\item (243)~Ida: identification number $2\,431\,010$, and
\item (243)~1~Dactyl,  Ida's satellite, identification number $2\,431\,011$.
\end{itemize}
The identification numbers for these asteroids were assigned using an older numbering convention now abandoned by the SPICE system. A conflict will arise between the new numbering system and the identification number for Ida if more than 431,010 asteroids are ever identified and cataloged. At that time NAIF will add another exception to the asteroid numbering system.

\end{document}